\documentclass[manuscript]{emulateapj}

\usepackage{amssymb}
\usepackage[export]{adjustbox}
\usepackage{amsmath}
\usepackage{booktabs}
\usepackage{hyperref}
\usepackage{multirow}
\usepackage{times}
\PassOptionsToPackage{hyphens}{url}\usepackage{hyperref}
\usepackage[hyphenbreaks]{breakurl}

\newcommand{\arcs}{\hbox{$^{\prime\prime}$}}

\newcommand{\ls}
{\mathrel{\hbox{\rlap{\hbox{\lower4pt\hbox{$\sim$}}}\hbox{$<$}}}}
\newcommand{\gs}
{\mathrel{\hbox{\rlap{\hbox{\lower4pt\hbox{$\sim$}}}\hbox{$>$}}}}

\begin{document}
\title{Resolving the X-ray obscuration in a low flux observation of the quasar PDS 456.}
\shorttitle{X-ray obscuration in PDS 456}
\shortauthors{Reeves et al.}
\author{J. N. Reeves\altaffilmark{1}, V. Braito\altaffilmark{1,2}, E. Nardini\altaffilmark{3}, F. Hamann\altaffilmark{4}, G. Chartas\altaffilmark{5}, A. P. Lobban\altaffilmark{6}, P. T. O'Brien\altaffilmark{7}, T. J. Turner\altaffilmark{8}} 
\altaffiltext{1}{Center for Space Science and Technology, 
University of Maryland Baltimore County, 1000 Hilltop Circle, Baltimore, MD 21250, USA; email jreeves@umbc.edu}
\altaffiltext{2}{INAF, Osservatorio Astronomico di Brera, Via Bianchi 46 I-23807 Merate (LC), Italy}
\altaffiltext{3}{INAF - - Osservatorio Astrofisico di Arcetri, Largo Enrico Fermi 5, I-50125 Firenze, Italy}
\altaffiltext{4}{Department of Physics \& Astronomy, University of California, Riverside, CA 92507, USA}
\altaffiltext{5}{Department of Physics \& Astronomy, College of Charleston, Charleston, SC 29424, USA}
\altaffiltext{6}{Astrophysics Group, School of Physical and Geographical Sciences, Keele 
University, Keele, Staffordshire, ST5 5BG, UK}
\altaffiltext{7}{Dept of Physics and Astronomy, University of Leicester, University Road, Leicester LE1 7RH, UK}
\altaffiltext{8}{Department of Physics, University of Maryland Baltimore County, 1000 Hilltop Circle, Baltimore, MD 21250, USA}

\begin{abstract}

Simultaneous {\it XMM-Newton}, {\it NuSTAR} and {\it HST} observations, performed in March 2017, of the nearby ($z=0.184$) luminous quasar PDS\,456 are presented. PDS\,456 had a low X-ray flux compared to past observations, where the 
first of the two new {\it XMM-Newton} observations occurred during a pronounced dip in the X-ray lightcurve. The broad-band 
X-ray spectrum is highly absorbed, attenuated by a soft X-ray absorber of column density 
$N_{\rm H}=6\times10^{22}$\,cm$^{-2}$. An increase in obscuration occurs during the dip, which may be due to an X-ray eclipse.
In addition, the persistent, fast Fe K outflow is present, with velocity components of $-0.25c$ and $-0.4c$. 
The soft absorber is less ionized ($\log\xi=3$) compared to the iron K outflow ($\log\xi=5$) and is outflowing with a velocity of approximately $-0.2c$. 
A soft X-ray excess is present below 1\,keV against the highly absorbed continuum and can be attributed to the re-emission from a wide angle wind. 
The complex X-ray absorption present in PDS\,456 suggests that the wind is inhomogeneous, whereby the soft X-ray absorber originates from denser clumps or filaments which may form further out along the outflow. 
In contrast to the X-ray observations, the simultaneous UV spectrum of PDS\,456 is largely unabsorbed, where only a very weak broad absorption trough is present bluewards of Ly$\alpha$, compared to a past observation in 2000 when the trough was significantly stronger. The relative weakness of the UV absorption may be due to the soft X-ray absorber being too highly ionized and almost transparent in the UV band. 
\end{abstract}










\section{Introduction}

Since their initial detection more than a decade ago \citep{Chartas02, Chartas03, Reeves03, Pounds03}, the presence of ultra fast outflows 
have been found to be increasingly common in Active Galactic Nuclei. They were discovered through the signature of highly blue-shifted absorption lines in the Fe K band in the X-ray spectra of AGN \citep{Tombesi10,Gofford13}. 
Given the large velocities (v$\sim 0.05-0.3c$)
and column densities ($\sim10^{23}$\,cm$^{-2}$ or higher) of these systems, 
their mass outflow rate can be comparable to the quasar accretion rate (up to several 
solar masses per year), while the mechanical power of the outflows 
can reach a significant fraction of the AGN bolometric luminosity \citep{PoundsReeves09,Tombesi13,Gofford15,Nardini15}. 
At high redshifts, such AGN winds could have provided the mechanical feedback
that controlled both the formation of stellar bulges and simultaneously self-regulated SMBH
growth, leading ultimately to the observed $M$--$\sigma$ relation for galaxies 
\citep{FerrareseMerritt00,Gebhardt00}.
Indeed evidence for the latter process has come from the discovery of massive, large (kpc) scale molecular outflows \citep{Feruglio10,Veilleux13,Cicone14}, which might be driven by the ultra fast outflows launched near to the black hole \citep{Tombesi15,Feruglio15,Feruglio17,Fiore17}.

At a redshift of $z=0.184$ and with a total bolometric luminosity of $\sim10^{47}$\,erg\,s$^{-1}$, PDS\,456 is the most luminous QSO in the local Universe \citep{Torres97, Simpson99, Reeves00} and is likely accreting near to the Eddington rate. It also hosts one of the best studied examples of an X-ray ultra fast outflow. The fast wind in PDS\,456 was first discovered in an {\it XMM-Newton} observation in 2001 \citep{Reeves03}, through the detection of deep blueshifted absorption troughs, both above 7\,keV in the iron K band and above 1\,keV in the iron L-shell band. 
Since this initial detection, the persistence of the ultra fast outflow in PDS\,456 became established through over a decade's worth of X-ray observations \citep{Reeves09,Behar10,Reeves14,Gofford14,Nardini15,Hagino15,Matzeu16,Matzeu17,Parker18}. It is also one of the fastest known winds, 
with the outflow velocity usually measured between $\sim0.25-0.3$\,c in all of the X-ray observations, with the exception of an additional faster ($\sim0.4c$) relativistic component 
which was recently revealed in a highly absorbed {\it NuSTAR} observation in 2017 \citep{Reeves18a}.
The wide-angle emission from the wind, as measured through the P Cygni profile at iron K \citep{Nardini15}, 
also implied that the wind subtends a large solid angle and that its mass outflow rate likely approaches the Eddington rate.

Both the X-ray continuum and the wind in PDS\,456 are highly variable. The column density of the iron K absorption can vary by up to an order of magnitude over a few days \citep{Gofford14,Matzeu16}, implying that the fast wind is likely located within a few hundred gravitational radii of the black hole. 
Recently, a positive correlation was discovered between the outflow velocity and the 
X-ray luminosity \citep{Matzeu17}, which suggests radiation plays a role in accelerating the initial wind. The soft X-ray spectrum of PDS\,456 can become strongly absorbed, for instance the {\it Suzaku} observations in 2013 caught PDS\,456 with a highly absorbed spectrum ($N_{\rm H}>10^{23}$\,cm$^{-2}$) at an extremely low flux \citep{Matzeu16}. The soft X-ray absorption may arise from a clumpy component of the wind, which is capable of partially covering the X-ray source. 
Indeed, from an analysis of the {\it XMM-Newton} Reflection Grating Spectrometer (RGS) observations of PDS\,456, \citet{Reeves16} showed that the soft X-ray spectrum exhibits blue-shifted absorption troughs, which may arise from a lower ionization 
component of the wind. Recent observations of other AGN are now also revealing soft X-ray components of ultra fast outflows, implying that the wind structure may be more complex than a simple homogeneous outflow \citep{Pounds03, Longinotti15, Reeves18b, Danehkar18,Pinto18}.

In contrast to the well studied X-ray wind in PDS\,456, its UV spectrum has been poorly studied. An initial observation with {\it HST} STIS in 
2000 revealed the presence a broad absorption trough, blueshifted by $-0.06c$ with respect to the broad Ly$\alpha$ emission line in the UV spectrum \citep{O'Brien05}. Furthermore, several of the broad UV emission lines, such as from C\,\textsc{iv}, were blueshifted by several thousand kilometers per second, suggesting the presence of a BLR scale wind component in PDS\,456.
A similar discovery was also recently made by \citet{Kriss18a}, who found a blue-shifted Ly$\alpha$ trough associated with the 
ultra fast outflow in the QSO, PG\,1211+143. In PDS\,456, \citet{Hamann18} suggested that the Ly$\alpha$ trough may instead be associated with a highly blue-shifted C\,\textsc{iv} absorption line, implying a UV outflow velocity of $-0.3c$, similar to the wind velocity measured in X-rays. 
Since then, only one other UV spectrum of PDS\,456 has been obtained, by {\it HST} COS in 2014, where the Ly$\alpha$ absorption trough appeared 
to be weaker compared to 2000 \citep{Hamann18}. However, neither of the 2000 or 2014 UV spectra had simultaneous X-ray coverage and thus the link between the X-ray and UV outflows in PDS\,456 is poorly understood.

\begin{deluxetable}{lcccc}
\tablecaption{X-ray Observation Log}
\tablewidth{0pt}
\tablehead{
& \colhead{NuSTAR} & \colhead{XMM OBS1} &  \colhead{XMM OBS2} & \colhead{Swift}}
\startdata
Sequence & 60201020002 & 0780690201 & 0780690301 & --\\
Start date & 2017/03/23 & 2017/03/23 & 2017/03/25 & 2017/03/23\\
Start time$^{a}$ & 05:31:09 & 18:56:04 & 05:58:10 & 11:30:57\\
End date & 2017/03/26 & 2017/03/24 & 2017/03/26 & 2017/04/10\\
End time$^{a}$ & 18:36:09 & 17:13:40 & 06:28:56 & 08:23:57\\
Duration$^{b}$ & 304.9\,ks & 79.3\,ks & 88.2\,ks & --
\enddata
\tablenotetext{a}{Observation start and end time in UT.}
\tablenotetext{b}{Observation duration in ks.}
\end{deluxetable}

\begin{deluxetable}{lcc}
\tablecaption{List of X-ray Exposures}
\tablewidth{0pt}
\tablehead{
\colhead{Detector} & \colhead{OBS 1} & \colhead{OBS2}}
\startdata
Exposures$^{a}$:-\\
XMM/pn & 39.6 & 64.9\\
XMM/MOS & 62.2 & 84.9\\
XMM/RGS & 72.0 & 88.0\\
NuSTAR & 38.2 & 47.2 \\
Count Rates$^{b}$:-\\
XMM/pn & $0.763\pm0.004$ & $1.174\pm0.004$\\
XMM/MOS & $0.420\pm0.003$ & $0.635\pm0.003$\\
XMM/RGS & $0.036\pm0.001$ & $0.058\pm0.001$\\
NuSTAR & $0.072\pm0.002$ & $0.096\pm0.002$
\enddata
\tablenotetext{a}{Net exposure times in ks after background screening and deadtime correction. Note that exposures are listed for an individual detector, e.g. MOS\,1 or RGS\,1.}
\tablenotetext{b}{Net source count rates in s\,$^{-1}$, after background substraction. Count rates are for MOS\,1+2, RGS\,1+2 and 
NuSTAR FPMA+FPMB combined respectively.}
\end{deluxetable}

In this paper we present a detailed analysis of a two recent simultaneous {\it XMM-Newton} and {\it NuSTAR} observations of PDS\,456, in March 2017.
These observations caught PDS\,456 in a low flux state, similar to the 2013 {\it Suzaku} observations, where a prominent decrease in flux was observed during the first of these two observations. In \citet{Reeves18a} (hereafter Paper I), we analysed the mean {\it NuSTAR} spectrum from the 2017 observations, which revealed a new ultra fast component of the Fe K outflow, with an outflow velocity of $\sim0.4c$, in addition to the persistent $\sim0.25c$ wind. Here, we present the broad-band X-ray spectral analysis, where the soft X-ray spectra are found to be strongly absorbed by a lower ionization component of the wind. In addition, a simultaneous {\it HST} COS observation was performed during the X-ray observations, in order to better understand the link between the X-ray and UV wind in PDS\,456. 

The structure of this paper is as follows; in Section 2 we describe the new observations and X-ray variability of PDS\,456, while in Section 3 we present the results of the broad-band X-ray spectral analysis from the simultaneous March 2017 observations. 
In Section 4, we present the multi-wavelength properties of PDS\,456, including the simultaneous 2017 {\it HST} spectrum and the overall 
Spectral Energy Distribution (SED). In Section 5, we discuss the properties of the soft X-ray absorber and its connection with the UV absorption in PDS\,456.

\begin{figure*}
\begin{center}
\rotatebox{0}{\includegraphics[width=14cm]{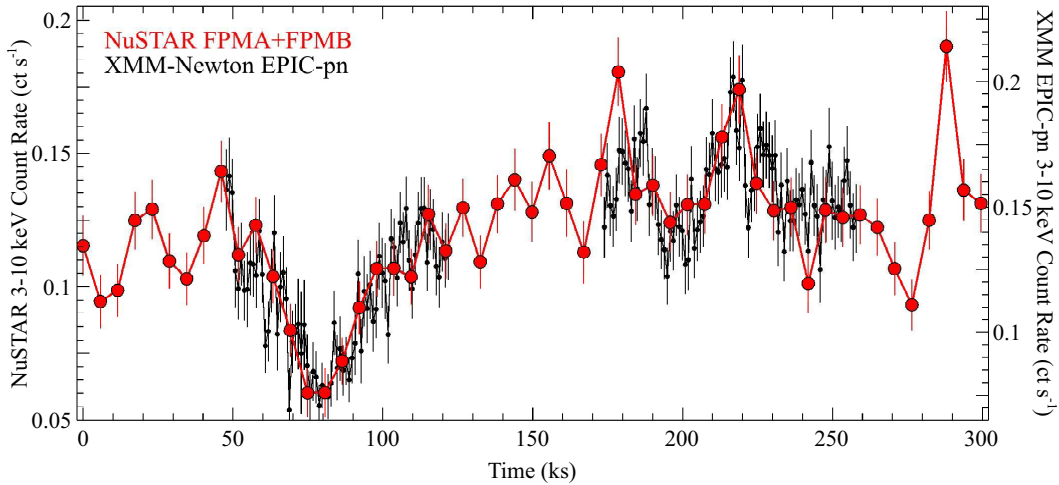}}
\end{center}
\caption{\small X-ray lightcurves of PDS 456 during the 2017 {\it XMM-Newton} and {\it NuSTAR} campaign. The background subtracted {\it NuSTAR} (red circles) and {\it XMM-Newton} (black points) lightcurves were extracted over their common 3--10\,keV band. 
Here the two {\it XMM-Newton} sequences correspond to OBS\,1 and OBS\,2 respectively.
Note that the first {\it XMM-Newton} observation (OBS\,1) coincided with a pronounced dip in the lightcurve.} 
\label{fig:lightcurves}
\end{figure*}

\begin{figure}
\begin{center}
\rotatebox{-90}{\includegraphics[height=8.7cm]{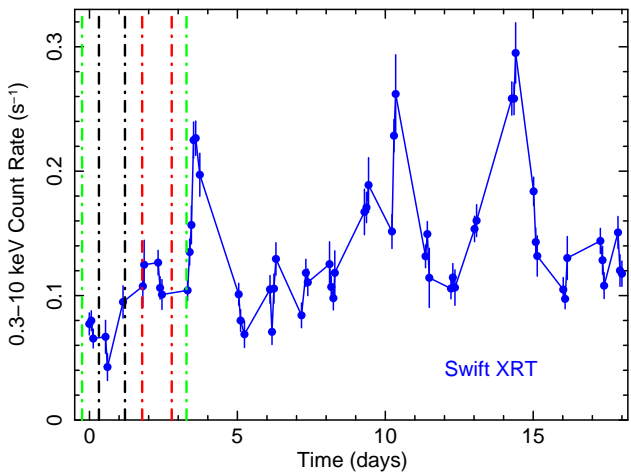}}
\end{center}
\caption{\small {\it Swift} daily monitoring of PDS 456, over the first 18 days of this campaign. The start and stop times of {\it NuSTAR} and {\it XMM-Newton} observations are denoted by dashed green ({\it NuSTAR}), black ({\it XMM-Newton} OBS\,1) and red ({\it XMM-Newton} OBS\,2) vertical lines (see Table\,1 for details). Note the low flux of the pointed observations, especially {\it XMM-Newton} OBS\,1, compared to later in the {\it Swift} campaign when pronounced X-ray flares are present.} 
\label{fig:Swift}
\end{figure}

\section{X-ray Observations and Data Reduction}

PDS\,456 was observed with {\it NuSTAR} \citep{Harrison13} from March 23--26, 2017, with a total duration of 305\,ks. This coincided with two simultaneous {\it XMM-Newton} observations, hereafter OBS\,1 and OBS\,2, each of approximately 80\,ks duration. 
The {\it XMM-Newton} observations were taken over two consecutive satellite orbits and with both in Large Window mode for EPIC-pn \citep{Struder01} and EPIC-MOS \citep{Turner01}; see Table\,1 for details of the observations. 
All data were processed using the \textsc{nustardas} v1.7.1, 
{\it XMM-Newton} \textsc{sas} v16.0 and \textsc{heasoft} v6.20 software.
{\it NuSTAR} source spectra were extracted using a 50\arcs\ circular region centered on the source and background from a 65\arcs\ circular region clear from stray light. 
{\it XMM-Newton} EPIC-pn and spectra were extracted from single and double events, using a 30\arcs\ radius source region and $2\times34\arcs$\ background regions on the same chip. 
Spectra from the RGS on-board 
{\it XMM-Newton} \citep{denHerder01} for PDS\,456 were extracted using the methods described 
in Reeves et al. (2016), where an analysis of the earlier archival {\it XMM-Newton} RGS spectra was presented. 
After checking for consistency, the RGS\,1 and RGS\,2 spectra were combined into a single RGS spectrum for each sequence, using the \textsc{sas} task \textsc{rgscombine}. 
The {\it XMM-Newton} Optical Monitor \citep{Mason01} or OM data were acquired in `Imaging Mode' and were processed using the \textsc{omichain} task within \textsc{sas}.  Source coordinates were verified using the \textsc{omsource} task and subsequent spectra were produced using \textsc{om2pha}.  Data were acquired with all six OM filters: V, B, U, UVW1, UVM2 and UVW2.  No source variability across the OM bandpass was detected within either observation and so we proceeded to analyze the time-averaged photometry.

Periods of high background flaring (due to soft Solar proton flares) were also removed from the two {\it XMM-Newton} observations. 
For the EPIC-pn detector, a threshold of 1~count\,s$^{-1}$ over the full CCD field over the 10-12\,keV band was used to reject events 
corresponding to high particle background. For OBS\,1, several background flares were present during the observation and as a result the net exposures were reduced to 39.6\,ks, 62.2\,ks and 72.0\,ks for the pn, MOS and RGS respectively, after removal of background flares and correcting for CCD deadtime. Note background flares are more severe for the pn CCD array than for the MOS CCDs and as a result the net exposure time was lower. During OBS\,2, the background level was more stable and as a result a higher net exposure time was obtained, of 64.9\,ks for EPIC-pn and 84.9\,ks for EPIC-MOS.
The subsequent net exposure times and count rates are listed in Table\,2, note that for the MOS and RGS the net count rates are for 
both detectors combined. After removal of the background flares, the background count rate is very low contributing to only 
2.6\% and 1.2\% of the total count rates over the 0.3-10\,keV band for the pn spectra of OBS\,1 and OBS\,2. Over the 3--10\,keV band the background rates are 5.8\% and 2.9\% of the total and have no effect on the net source spectrum over the iron K band.

Figure\,1 shows the EPIC-pn X-ray lightcurves of the two {\it XMM-Newton} observations, superimposed on the lightcurve from the whole {\it NuSTAR} observation. The lightcurves from both satellites were extracted over the common 3--10\,keV band and were binned into orbital (5814\,s) bins for the {\it NuSTAR} observation. PDS\,456 shows pronounced X-ray variability during these observations.
In particular, {\it XMM-Newton} OBS\,1 coincided with a pronounced dip in the source count rate at $\sim80$\,ks into the {\it NuSTAR} observation. In addition to the {\it NuSTAR} observations, PDS\,456 was monitored daily with {\it Swift} for 18 days during this campaign, with the {\it Swift} monitoring starting just after the beginning of the 2017 {\it NuSTAR} observation. Figure\,2 shows the {\it Swift} XRT 
lightcurve over the 0.3-10\,keV band, with the start and stop times of the {\it NuSTAR} and the two {\it XMM-Newton} observations 
shown for comparison. From the {\it Swift} lightcurve, it is apparent that the {\it XMM-Newton} observations occurred during a relatively low period of X-ray flux during this campaign, which is especially the case for OBS\,1. Note that the details of the {\it Swift} monitoring will be presented in a subsequent paper, where after the initial daily monitoring, PDS\,456 was monitored weekly for a further 6 months and which also shows pronounced 
long-term variability.

The total net exposure time of the whole {\it NuSTAR} observation is 157\,ks.
The mean {\it NuSTAR} spectrum 
was presented earlier in Paper I, where the presence of two velocity components of the fast outflow (with velocities $0.25c$ and $0.43c$) was revealed. 
In this paper, the main focus is on the broad-band analysis between the {\it XMM-Newton} and {\it NuSTAR} observations, 
using simultaneous exposures in the soft and hard X-ray bands.
As a result of the pronounced X-ray 
variability, spectra from the {\it NuSTAR} FPMA and FPMB detectors were extracted only during the 
time intervals corresponding to the start and stop times of the {\it XMM-Newton} OBS\,1 and OBS\,2 exposures (see Table 1), so that the simultaneous broad-band {\it XMM-Newton} and {\it NuSTAR} spectra could be analyzed for each sequence. Subsequently the net {\it NuSTAR} exposures obtained for each sequence were 38.2\,ks and 47.2\,ks, for the OBS\,1 and OBS\,2 intervals respectively. The {\it NuSTAR} spectra exclude time periods through and immediately after the passage of the satellite through the South Atlantic Anomaly. 
Furthermore, as the spectra obtained for the FPMA and FPMB modules were consistent, these were combined into single {\it NuSTAR} spectra coincident with each of the OBS\,1 and OBS\,2 {\it XMM-Newton} sequences. 
All spectra are binned to at least 50 counts per bin to allow the use of $\chi^{2}$ minimization. 
Outflow velocities are given in the rest-frame of PDS\,456 at $z=0.184$, 
after correcting for relativistic Doppler shifts along the line of sight. Note that fluxes are stated without correcting for either Galactic or intrinsic absorption, whereas luminosities are corrected for absorption and quoted in the rest-frame. Errors are quoted at 90\% confidence for one interesting parameter (or $\Delta\chi^2=2.7$).

\begin{figure*}
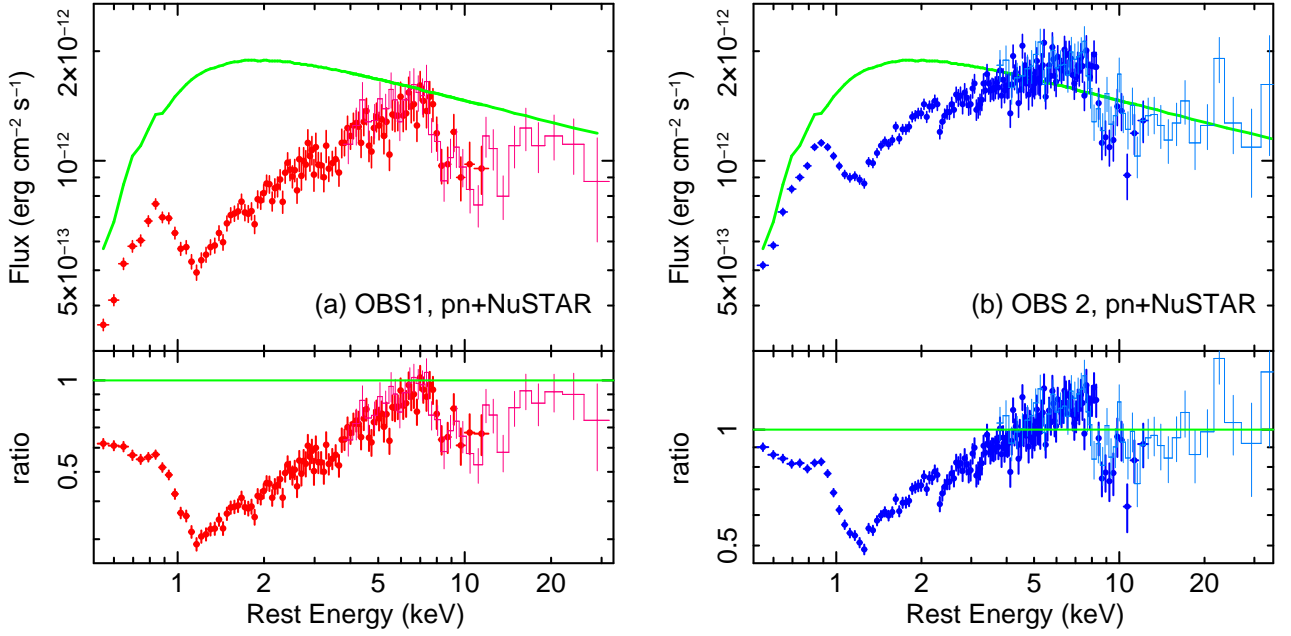

\begin{center}
\rotatebox{-90}{\includegraphics[height=8.7cm]{f3a.eps}}
\rotatebox{-90}{\includegraphics[height=8.7cm]{f3b.eps}}
\end{center}
\caption{\small Simultaneous {\it XMM-Newton}/pn and {\it NuSTAR} X-ray spectra from the 2017 observations of PDS\,456. 
The first {\it XMM-Newton} (OBS\,1) sequence is shown in red (panel (a)) and the OBS\,2 sequence in blue (panel (b)), while simultaneous {\it NuSTAR} spectra (shown in lighter shading) were extracted using identical 
time intervals as per the {\it XMM-Newton} observations; see Table\,1 for details of the exposures. 
A power-law continuum of photon index $\Gamma=2.3$, modified by Galactic absorption only, is shown as a solid green line, which was normalized to OBS\,2 above 4\,keV. The lower panels of each plot show the data/model residuals compared to the power-law 
continuum.
Both the OBS\,1 and OBS\,2 spectra are strongly modified by excess absorption, as seen by the large deficit of counts below 4\,keV, 
which is especially the case for the OBS\,1 spectrum, which it is at a lower flux overall.  A minimum of counts occurs near 1.2\,keV, 
above the iron L-shell band.
The residuals present above 8 keV are due to the iron K absorption, as described in paper I.
Note the spectra have been converted to flux ($\nu F_{\nu}$) units by unfolding the count rate spectra against a 
$\Gamma=2$ continuum only.} 
\label{fig:spectra}
\end{figure*}

\section{Broad-Band Spectral Analysis}

In this section, we concentrate on the broad-band X-ray spectral analysis, using the simultaneous {\it XMM-Newton} EPIC-pn and 
{\it NuSTAR} spectra taken during the OBS\,1 and OBS\,2 intervals. In particular, this 
provides an opportunity to study the broad-band X-ray spectrum of PDS\,456 during an unusually low X-ray flux observation.
Indeed the 2-10\,keV flux derived during OBS\,1 was $1.8\times10^{-12}$\,erg\,cm$^{-2}$\,s$^{-1}$, 
compared to the range of $2.4-7.4\times10^{-12}$\,erg\,cm$^{-2}$\,s$^{-1}$ observed in the 
previous joint {\it XMM-Newton} and {\it NuSTAR} observations (OBS\, A--E) in 2013--14 analyzed by \citet{Nardini15} and 
\citet{Matzeu17}.  
The pn and {\it NuSTAR} FPMA+FPMB spectra were fitted jointly, using the 0.3--10\,keV band for the pn and the 3.5--30\,keV band 
for {\it NuSTAR}; note for the latter the spectra become background dominated above 30\,keV and data above this energy 
were excluded from the spectral analysis.
A multiplicative constant of $1.05\pm0.03$ is included between the {\it NuSTAR} and pn spectra to allow for any difference in normalization through cross calibration and was allowed to vary as a free parameter in the modeling.
We included a component of Galactic absorption in the spectral fitting, using the \textsc{tbabs} model of \citet{Wilms00}, 
where for PDS\,456, the column density is expected to be $N_{\rm H}=2.4\times10^{21}$\,cm$^{-2}$ based on 21\,cm measurements \citep{Kalberla05}. Solar abundances of \citet{GrevesseSauval98} were used throughout.

\subsection{The Overall Spectral Form}

Figure~\ref{fig:spectra} shows the simultaneous pn and {\it NuSTAR} spectra from OBS\,1 and OBS\,2, compared to a $\Gamma=2.3$ 
powerlaw fitted to the OBS\,2 spectrum above 4\,keV for comparison and modified by Galactic absorption. This is close to the best-fit photon index found in the hard X-ray analysis in paper I, as well as from historical spectra of PDS\,456. A large deficit of counts is apparent against this simple powerlaw continuum in the soft X-ray band, with a minimum in the spectra occurring near to 1.2\,keV, blue-wards of the iron L-shell band.  This is especially the case for OBS\,1, where the observation occurred during the pronounced dip in the lightcurve and it appears at first glance that the spectrum is strongly obscured. Strong absorption is also present between 8-12\,keV, blue-wards of the iron K-shell band, which again is very prominent in the OBS\,1 spectrum. Note that similar broad-band residuals are also 
present from the EPIC-MOS spectra, in the form of absorption troughs in both the iron L and iron K band.  No neutral reflection component, associated with a distant reprocessor, has been included in the modeling, as there is no neutral iron K$\alpha$ line nor Compton hump present in the X-ray spectrum, consistent with what has been found from previous observations \citep{Reeves03,Reeves09,Nardini15}. The formal upper limit on the equivalent width of a narrow 6.4\,keV 
K$\alpha$ line is $<45$\,eV, measured against the OBS\,1 continuum.

We constructed an initial baseline model in order to account for the broad band spectra. This consisted of a power-law continuum, absorbed by a simple 
one-zone photoionized absorber which is allowed to partially cover the X-ray source. Thus the model is in the phenomenological form of:-\\

${\rm tbabs} \times [(1-f)\times{\rm pow} + f\times{\rm xstar}_{\rm pcov}\times{\rm pow}]$\\

where $f$ is the fraction of the power-law continuum (${\rm pow}$) 
that passes through the ionized absorber (denoted above by ${\rm xstar}_{\rm pcov})$ and 
thus the fraction $1-f$ is unattenuated. 
Here \textsc{tbabs} represents the neutral Galactic photoelectric absorber, which 
absorbs the entire spectrum. A multiplicative grid of absorption models, generated by the \textsc{xstar} photoionization 
code \citep{Kallman04}, was used to model the absorption. 
Here we adopt the same absorption model grids used in \citet{Nardini15}, 
where the mean optical to X-ray SED of PDS\,456 was used as the input continuum, which has 
an ionizing (1--1\,000\,Ryd) luminosity of $L_{\rm ion}=5\times10^{46}$\,erg\,s$^{-1}$. A turbulence velocity width of 15\,000\,km\,s$^{-1}$ was used, which accounts for the velocity broadening of the Fe K absorption features \citep{Nardini15,Reeves18a}. 

The column density, outflow velocity, ionization parameter and covering factor of the photoionized absorber were allowed to vary 
for each observation during the modeling, as was the normalization of the power-law continuum.
The best-fit absorption parameters derived from this model consist of, for OBS\,1; $N_{\rm H}=1.20^{+0.04}_{-0.07}\times10^{23}$\,cm$^{-2}$, an ionization parameter of $\log\xi=3.42\pm0.02$\,erg\,cm\,s$^{-1}$, a covering fraction of 
$f=0.77\pm0.02$ and an outflow velocity of $-0.239\pm0.006c$. 
For OBS\,2, the spectrum is slightly less 
absorbed, with a lower column and covering fraction of 
$N_{\rm H}=0.94^{+0.05}_{-0.05}\times10^{23}$\,cm$^{-2}$ and $f=0.70\pm0.02$ respectively, while the ionization parameter and outflow velocity are consistent between the two epochs. The continuum photon index is $\Gamma=2.43\pm0.03$, determined by the slope of the spectrum in the highest energy band.
Note that the high outflow velocity of the soft X-ray absorber is driven by the need to fit the broad absorption trough above 1\,keV with a blend of transitions from iron L; the fit statistic worsened considerably from $\chi_\nu^{2}=1139/663$ to 
$\chi_\nu^{2}=1892/664$ if the outflow velocity was assumed to be zero, leaving large negative residuals between $1-1.5$\,keV.
The velocity is also consistent with the velocity of the low ionization partial covering absorber derived by \citet{Matzeu16}, obtained during a similar low flux observation in 2013 with
{\it Suzaku}. It is also in agreement with what was inferred from the archival analysis of the previous {\it XMM-Newton} RGS spectra, 
from the presence of broad absorption troughs in the soft X-ray band \citep{Reeves16}. 

\begin{figure}
\begin{center}
\rotatebox{-90}{\includegraphics[height=8.7cm]{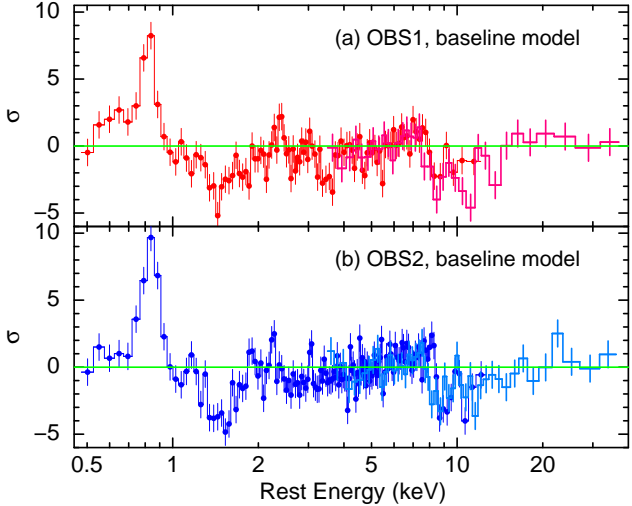}}
\end{center}
\caption{\small Residuals against the initial baseline partial covering model described in Section\,3; panel (a) shows OBS\,1 and panel (b) OBS\,2. The residuals present in the iron K-band are due to the two-component ultra fast outflow, with outflow velocities of $-0.25c$ and $-0.43c$ as described in paper I. After modeling the spectra with the baseline model, strong residuals are still present in the soft X-ray band. An excess of counts is observed near to 0.8 keV, while additional absorption is present above 1 keV; these remaining features may be associated with emission and absorption from L-shell iron originating from the wind, in addition to the iron K band absorption above 8\,keV.} 
\label{fig:residuals}
\end{figure}

Nonetheless, despite the fact that the absorption model is able to reproduce the overall shape of the spectra through the large opacity and covering fraction of the absorber, the fit statistic is very poor, with $\chi_\nu^{2}=1139/663$, which is rejected with a null hypothesis probability of $P=1.2\times10^{-27}$.  
Strong residuals are still apparent in the data, compared to this baseline absorption model, as is shown in Figure~\ref{fig:residuals}.
In particular, a strong excess in emission is seen in both spectra between $0.8-0.9$\,keV, 
which can be parameterized by a broad Gaussian profile, with an equivalent width of $85\pm15$\,eV and a 
velocity width of $\sigma_{\rm v}=14000\pm5000$\,km\,s$^{-1}$, at a centroid energy of $0.88\pm0.01$\,keV. 
Furthermore, part of the broad absorption trough 
is missed between 1--2\,keV, while the absorber is not able to model the iron K band absorption seen between $8-12$\,keV.
This suggests that a single zone of absorption is not fully able to account for all of the features present in the data and that gas covering a wider range of ionization may be present. 
The relatively low ionization ($\log\xi=3.42\pm0.05$\,erg\,cm\,s$^{-1}$) of this absorber is driven 
primarily by the requirement to model the strong soft X-ray absorption present in the data, which likely originates from more 
moderately ionized iron (Fe\,\textsc{xvii-xxiv}) and from lighter elements, but which leaves the high energy absorption from the He/H-like K-shell lines of iron unmodeled as a result.

\begin{figure*}
\begin{center}
\rotatebox{-90}{\includegraphics[height=13cm]{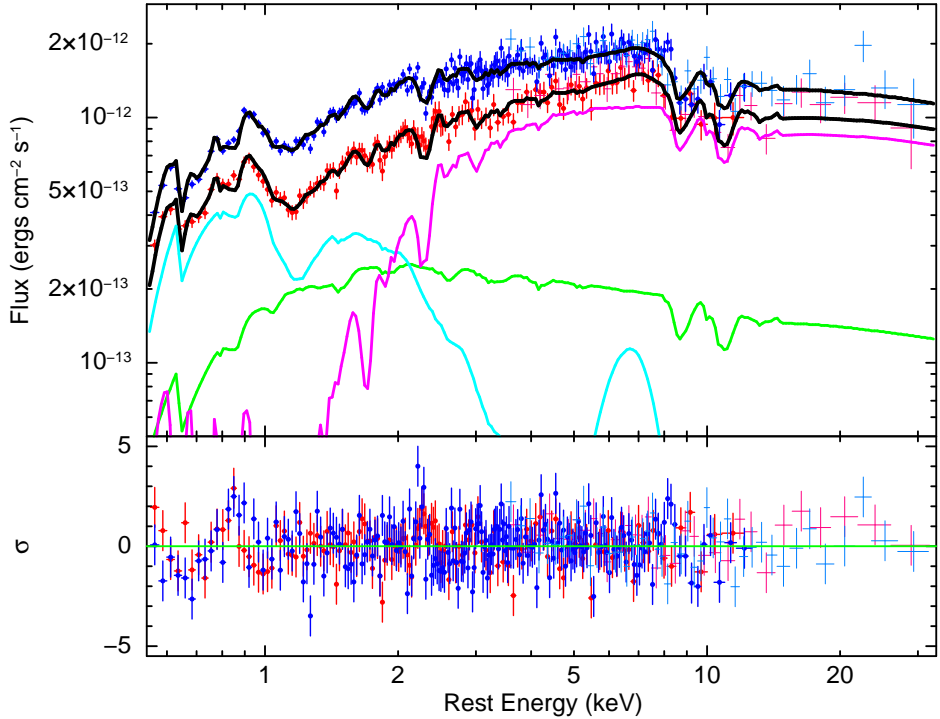}}
\end{center}
\caption{\small Fluxed spectra from the two observations, with the best-fit \textsc{xstar} model super-imposed. 
The red and blue spectra correspond to OBS\,1 and OBS\,2, respectively.
The solid black line shows the total model for both observations, while the other solid lines illustrate the model deconstruction for OBS\,1 only. 
Here, the green line represents the powerlaw continuum 
absorbed through the two high ionization wind zones only (zones 1a and 1b), while the magenta line is the continuum component that is additionally absorbed through the lower ionization absorber (zone 2), 
which produces the opacity present at soft X-rays below 3\,keV. The cyan line then represents the soft X-ray excess emission, from a photo-ionized emitter with a velocity broadening of $\sigma=0.1c$. The latter is able to account for the line emission below 1\,keV, as well as the broad emission at iron K.} 
\label{fig:xstar}
\end{figure*}

\subsection{The Best-fit Broad-Band Model}

Thus, following the approach of paper I, we added two additional (higher ionization) zones of absorption fully covering the continuum, in order to account for the blue-shifted absorption features seen in the iron K band spectrum. 
This was modeled in paper I, on the basis of the mean 2017 {\it NuSTAR} spectrum, by two outflowing absorbers, with outflow velocities of $-0.25c$ and $-0.43c$, originating from the blueshifted resonance ($1s\rightarrow2p$) 
absorption lines of He/H-like iron.
The soft X-ray partial covering absorber is retained, as in the above model construction.
In addition we also include emission from photoionized gas, to account for the 
excess emission observed below 1\,keV, as is seen in the residuals in Fig\,4. 
Furthermore, in order to account for any velocity broadening of the emission, we followed the same methodology employed in Nardini et al. (2015) and 
used an additive \textsc{xstar} photoionized emission table generated for PDS\,456, which is convolved with a Gaussian function.
Thus the form of this final model is:-\\

\noindent ${\rm tbabs} \times {\rm xstar}_{\rm Fe K} \times [(1-f) \times {\rm pow} + f \times {\rm xstar_{\rm pcov}} \times {\rm pow} + {\rm gauss}\otimes{\rm emiss}]$\\

where ${\rm xstar}_{\rm Fe K}$ denotes the iron K absorbers and ${\rm gauss}\otimes{\rm emiss}$ represents the 
photoionized emission convolved with a Gaussian profile. All the emission components are absorbed 
by a Galactic component of absorption, via the \textsc{tbabs} model, as above.

\begin{figure}
\begin{center}
\rotatebox{-90}{\includegraphics[height=8.7cm]{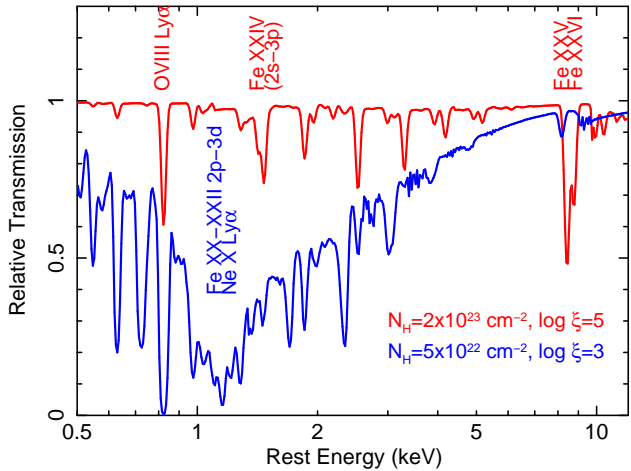}}
\end{center}
\caption{\small Relative transmission through two zones of absorption, representative of the absorption seen towards PDS\,456 in these observations. Both zones are blueshifted by $-0.25c$ and are corrected for Galactic absorption. The high ionization absorber (shown in red) contributes towards the 
Fe K absorption, as seen through the $1s\rightarrow2p$ lines of Fe\,\textsc{xxv} and Fe\,\textsc{xxvi} observed near to 8.5\,keV. The lower ionization zone (shown in blue) has much greater opacity in the soft X-ray band and in particular can reproduce the 
broad absorption trough seen near to 1.2\,keV in these observations, as well as the strong spectral curvature. Both zones also 
produce blueshifted O\,\textsc{viii} Ly$\alpha$ absorption near to 0.8 keV, as is also detected in the RGS spectrum. 
The simulation also illustrates the wealth of wind features that may be observed in future X-ray calorimeter spectra, 
from {\it XARM} and {\it Athena}.}  
\label{fig:transmission}
\end{figure}

Overall this model produces an acceptable fit to the OBS\,1 and OBS\,2 spectra, with a fit statistic of $\chi^2=690/650$, significantly improved compared 
to the above model with just a single absorber zone and no additional emission. Figure~\ref{fig:xstar} shows the resulting best-fit to both datasets, where no strong residuals now remain compared to the model. The relative contribution of the emission components 
are shown as fitted to OBS\,1; here the unabsorbed powerlaw (green line) is absorbed only by the high ionization (and low opacity) Fe K absorber, while the magenta line shows the absorbed powerlaw which is additionally covered by the lower ionization 
(high opacity) absorber in the soft band. 
The much higher normalization of the absorbed power-law component, compared to the unabsorbed power-law,  
can be seen by their relative contributions above 10\,keV in Figure~\ref{fig:xstar}, indicating that the covering fraction of the lower ionization soft X-ray absorber is high, covering nearly 90\% of the continuum in the case of OBS\,1 (see Table\,3 for details of the model parameters). The soft emission component (Figure~\ref{fig:xstar}, cyan line) makes a strong contribution towards the soft X-ray continuum and is able to model the excess line emission below 1\,keV, as well as some broadened line emission in the iron K band.
 Its ionization is $\log\xi=3.43\pm0.05$, similar to that of the partial covering absorber and the emitter produces a blend of 
emission lines from L-shell Fe (Fe\,\textsc{xvii-xiv}) as well as from He/H-like Ne and O.
A large velocity broadening is required for this emission component, with a Gaussian width of $\sigma=0.80\pm0.05$\,keV measured at 6 keV.  As will be discussed in Section\,3.5, this may represent the broadened emission component from a wide angle wind.  

Two outflowing absorption zones are required at Fe K (denoted as ${\rm xstar}_{\rm Fe K}$ above), one slower zone with $v_{\rm out}=-0.25\pm0.02c$ and the faster zone with $v_{\rm out}=-0.43\pm0.02c$; see Table\,3 for details.
Both outflowing zones at Fe K are highly significant, their addition to the baseline model improved the 
fit by $\Delta\chi^2=-72.5$ and $\Delta\chi^2=-36.4$ for the slower and faster zones (hereafter zones 1a and 1b) respectively. 
As can be seen in Figure~\ref{fig:xstar}, these two iron K absorbers are able to account for the absorption observed between 
$8-12$\,keV in both of the OBS\,1 and OBS\,2 spectra, with the faster zone reproducing the higher energy absorption trough, 
as was described in depth in paper I. The iron K zones are much more highly ionized than what is required to model the soft X-ray spectrum, with a ionization parameter of $\log\xi=4.9\pm0.1$ for the slower zone 1a and a lower limit on the ionization of $\log\xi>5.6$ for the faster zone 1b, which renders this latter component to be completely transparent in the soft X-ray band.

In contrast, the soft X-ray absorber is of much lower ionization (Table\,3, zone 2), with $\log\xi=3.05\pm0.05$ and reproduces the strong spectral curvature seen below 3\,keV in both spectra, as well as the broad absorption trough above 1\,keV. Figure~\ref{fig:transmission} shows the relative opacity (compared to a $\Gamma=2$ continuum) of both this lower ionization zone, with $\log\xi=3$ and $N_{\rm H}=5\times10^{22}$\,cm$^{-2}$, compared to the iron K zone absorber 
(zone 1a above), which has an ionization parameter two orders of magnitude higher with $\log\xi=5$ and $N_{\rm H}=2\times10^{23}$\,cm$^{-2}$. Despite its lower column, the low ionization absorber produces much more opacity in the soft X-ray band, as well as a broad absorption trough at $\sim1.2$\,keV, 
as is observed in both of the {\it XMM-Newton} spectra, which is primarily due to a blend of L-shell 
transitions of iron as well as absorption from Ne\,\textsc{ix-x}. 
The outflow velocity of the soft X-ray absorber is $-0.23\pm0.02c$, similar to the zone 1a Fe K absorber and also similar to that inferred 
in the analysis of the previous soft X-ray spectra of PDS\,456 in Reeves et al. (2016).
The high ionization absorber produces mainly the strong resonance absorption in the Fe K band, 
through the blueshifted $1s\rightarrow2p$ transitions of He and H-like iron (Fe\,\textsc{xxv} and Fe\,\textsc{xxvi}). 
In contrast, this zone produces relatively little opacity 
in the soft band, compared to the lower ionization absorber. Note that both zones do predict a strong blue-shifted O\,\textsc{viii} Ly$\alpha$ absorption line in the soft X-ray band, which, as we will show in Section\,3.4, is detected in the higher resolution 
{\it XMM-Newton} RGS spectrum.

\begin{figure*}
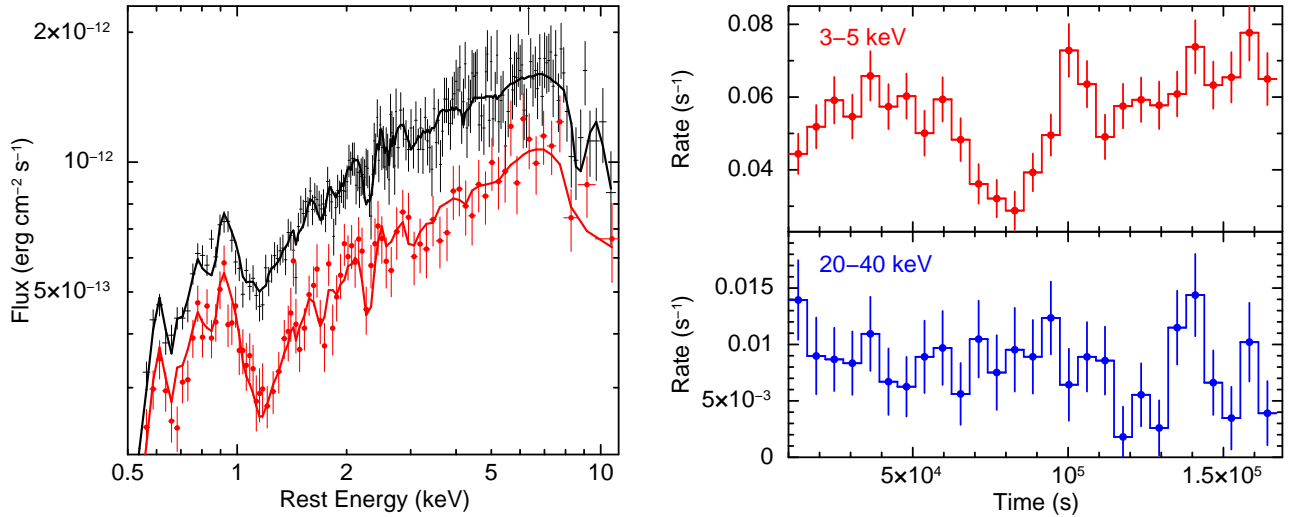

\begin{center}
\rotatebox{-90}{\includegraphics[height=8.7cm]{f7a.eps}}
\rotatebox{-90}{\includegraphics[height=8.7cm]{f7b.eps}}
\end{center}
\caption{\small Left panel:- EPIC-pn spectra of PDS\,456 during OBS\,1, selected during the minimum in flux during the lightcurve dip (red circles) versus the remainder of the observation (black crosses). The dip spectrum is lower in flux, has a strong absorption trough above 1\,keV and can be fitted by an increase in covering fraction of the soft X-ray absorber to $>97$\% covering. The solid lines show the best fit variable covering model overlaid on the two spectra. Right panel:- a portion of the net {\it NuSTAR} lightcurves showing the source variability during the dip in the lightcurve over the 3--5\,keV and 20--40\,keV bands. The pronounced dip is only seen over the softer 3-5\,keV band, but not in the 20--40\,keV band. Thus the dip is most likely due to an increase in obscuration affecting the softer X-ray band, but which has little effect on the hard X-ray continuum from 20-40\,keV.} 
\label{fig:dip}
\end{figure*}

\begin{deluxetable*}{lccc}
\tablecaption{Spectral Parameters for the 2017 {\it XMM-Newton} and {\it NuSTAR} Observations.}
\tablewidth{0pt}
\tablehead{\colhead{Parameter} & \colhead{OBS\,1} & \colhead{OBS\,2} & \colhead{OBS\,1 dip}}
\startdata
Fe K absorber, zone 1a:-\\
$N_{\rm H}$$^{a}$ & $2.15\pm0.45$ & $1.50^{+0.35}_{-0.25}$ & $2.15^f$\\
$\log\xi$$^{b}$ & $4.9\pm0.1$ & $4.9^t$ & $4.9^f$\\
$v/c$  & $-0.247^{+0.017}_{-0.012}$ & $-0.259^{+0.010}_{-0.013}$ & $-0.247^f$\\
\hline
Fe K absorber, zone 1b:-\\
$N_{\rm H}$$^{a}$ & $7.3^{+2.3}_{-2.2}$ & $3.1^{+1.5}_{-1.3}$ & $7.3^f$\\
$\log\xi$$^{b}$ & $5.8^{+1.2p}_{-0.2}$ & $5.8^t$ & $5.8^f$\\
$v/c$  & $-0.43\pm0.02$ & $-0.43^{t}$ & $0.43^f$\\
\hline
Soft X-ray absorber, zone 2:-\\
$N_{\rm H}$$^{a}$ & $0.66^{+0.04}_{-0.05}$ & $0.54\pm0.04$ & $0.62\pm0.05$\\
$\log\xi$$^{b}$ & $3.05\pm0.05$ & $3.05^t$ & $3.0\pm0.1$\\
$v/c$  & $-0.23\pm0.01$ & $-0.23^t$ & $-0.23^{f}$\\
$f^c$ & $0.87\pm0.03$ & $0.73\pm0.03$ & $>0.97$\\
\hline
Soft X-ray Emission:-\\
$N_{\rm H}$$^{a}$ & $0.66^f$ & $0.54^f$ & $0.62^t$\\
$\log\xi$$^{b}$ & $3.43\pm0.05$ & $3.43^t$ & $3.43^f$ \\
$\kappa_{\rm xstar}$$^{d}$ & $0.68\pm0.06\times10^{-3}$ & $0.82\pm0.07\times10^{-3}$ & $0.68^f$ \\
$\Omega/4\pi$$^{e}$ & $0.72\pm0.07$ & $0.86\pm0.09$ & --\\
\hline
Continuum:-\\
$\Gamma$ & $2.29\pm0.04$ & $2.29^{t}$ & $2.29^f$\\
$N_{\rm ABS}$$^{g}$ & $1.36\pm0.10$ & $1.36^t$ & $1.08\pm0.06$\\
$N_{\rm UNABS}$$^{g}$ & $0.20\pm0.03$ & $0.50\pm0.04$ & $<0.036$\\ 
$F_{\rm 2-10\,keV}$$^{h}$ & 1.83 & 2.55 & 1.20\\
$L_{\rm 2-10\,keV}$$^{i}$ & 2.71 & 3.44 & 1.84 \\
$L_{\rm 7-30\,keV}$$^{i}$ & 1.69 & 2.16 & --\\
\enddata
\tablenotetext{a}{Units of column density $\times10^{23}$\,cm$^{-2}$. Note for the Fe K zone 1b, the $N_{\rm H}$ has been calculated for a fixed ionization parameter of $\log\xi=5.8$.}
\tablenotetext{b}{Ionization parameter (where $\xi=L/nR^{2}$) in units of erg\,cm\,s$^{-1}$.}
\tablenotetext{c}{Covering fraction of absorber, where $f=1$ for a fully covering absorber.}
\tablenotetext{d}{Measured normalization of \textsc{xstar} emission component, where $\kappa_{\rm xstar}= f_{\rm cov} L_{38}/D_{\rm kpc}^2$ and $f_{\rm cov}$ is the emitter covering fraction ($f=\Omega/4\pi$), $L_{38}$ is the 1--1000\,Ryd ionizing luminosity in units of $10^{38}$\,erg\,s$^{-1}$ and $D_{\rm kpc}$ is the distance to PDS\,456 in units of kpc.}
\tablenotetext{e}{Solid angle of the emission component, derived from the measured normalization of the emitter. See Section\,3.5 for details.}
\tablenotetext{f}{Denotes parameter is fixed.}
\tablenotetext{g}{Relative normalizations of the absorbed and unabsorbed power-law continuum components. Normalizations are in flux units of $\times10^{-3}$\,photons\,cm$^{-2}$\,s$^{-1}$\,keV$^{-1}$ at 1\,keV.}
\tablenotetext{h}{Observed 2--10\,keV flux, not corrected for absorption, in units of $\times10^{-12}$\,erg\,cm$^{-2}$\,s$^{-1}$.}
\tablenotetext{i}{Intrinsic (absorption corrected) 2--10\,keV (or 7-30\,keV) luminosity, in units of $\times10^{44}$\,erg\,s$^{-1}$.}
\tablenotetext{p}{Denotes parameter error is pegged to the highest possible value in the model}
\tablenotetext{t}{Denotes parameter is tied between observations.}
\label{tab:xstar}
\end{deluxetable*}

\subsection{Spectral Variability}

Much of the spectral variability between the OBS\,1 and OBS\,2 spectra appears to be reproduced by the variability of the 
low ionization partial covering absorber (zone 2), with the spectrum somewhat less absorbed in OBS\,2 vs OBS\,1, although the former still requires substantial absorption. The column density and covering fraction of zone 2 for OBS\,1 
are $N_{\rm H}=6.6^{+0.4}_{-0.5}\times10^{22}$\,cm$^{-2}$ and $f=0.87\pm0.03$, which then decreases to $N_{\rm H}=5.4\pm0.4\times10^{22}$\,cm$^{-2}$ and $f=0.73\pm0.03$ during OBS\,2. Note that this change 
is primarily driven by a change in the covering fraction, the fit statistic subsequently worsens by $\Delta\chi^2=52$ for $\Delta\nu=1$ 
if the covering fraction is forced to be identical between both datasets.

Indeed the major difference between the two spectra is seen through the normalization of the unabsorbed power-law 
component; this more than doubles in flux from OBS\,1 to OBS\,2 from 
$N_{\rm UNABS} = 2.0\pm0.3\times10^{-4}$\,photons\,cm$^{-2}$\,s$^{-1}$\,keV\,$^{-1}$ (at 1\,keV) to $N_{\rm UNABS}=5.0\pm0.4\times10^{-4}$\,photons\,cm$^{-2}$\,s$^{-1}$\,keV\,$^{-1}$, while the normalization of the absorbed power-law component remains unchanged within the errors and thus was subsequently tied between the observations (see Table\,3 for details). 
Alternatively, the fraction of the power-law continuum ($1-f$) which emerges unattenuated by the soft X-ray absorber 
increases from $1-f = 0.13\pm0.03$ to $1-f=0.27\pm0.03$ from OBS\,1 to OBS\,2. 

\subsubsection{Variability around the Lightcurve dip}

To further test the change in obscuration around OBS\,1 and whether the dip in the lightcurve is due to variable absorption, we isolated the pn spectrum of the dip in the lightcurve, centered within $\pm10$\,ks of the minimum in flux seen in Figure~\ref{fig:lightcurves}. For comparison, we then extracted the pn spectrum of the remaining OBS\,1 pn exposure, minus the portion around the dip. This resulted in a net exposure of 12\,ks in the dip and 28\,ks for the non-dip portion of OBS\,1. The resulting spectra are shown in Figure~\ref{fig:dip} (left panel), where the dip spectrum is clearly fainter than the non-dip portion, while the absorption trough just above 1\,keV increased in depth during the dip and the spectrum became harder. 

We applied the above partial covering model to both spectra, varying only the covering fraction of the soft X-ray absorber (via the relative normalizations of the obscured and unobscured power-law components). The overall column density, ionization parameter and outflow velocity of the soft X-ray absorber was also assumed to remain constant across OBS\,1. The highly ionized iron K absorption (zones 1a and 1b) was fixed to the mean values found for OBS\,1, as there is not enough signal in the short dip spectrum to determine whether the Fe K profile varies or not. Similarly the soft X-ray emission component was  assumed to remain constant. 

The variations between the dip and non-dip spectra can be well modeled by a change in the covering fraction 
of the soft X-ray absorber and the overall fit statistic is good, with $\chi_{\nu}^2=269.2/276$. 
Overall the soft X-ray (0.5--2\,keV) flux dropped by 50\% from $F_{\rm 0.5-2.0 keV}=9.1\times10^{-13}$\,erg\,cm$^{-2}$\,s$^{-1}$ for the non-dip portion to 
$F_{\rm 0.5-2.0 keV}=5.7\times10^{-13}$\,erg\,cm$^{-2}$\,s$^{-1}$ during the dip.
The drop in flux can be accounted for by a decrease in normalization of the unobscured power-law to $N_{\rm UNABS}<3.6\times10^{-5}$\,photons\,cm$^{-2}$\,s$^{-1}$, formally consistent with zero. 
In comparison, the obscured power-law dominates the spectrum, where $N_{\rm ABS}=1.08\pm0.06 \times10^{-3}$\,photons\,cm$^{-2}$\,s$^{-1}$; see Table\,3 for details of the OBS\,1 dip parameters. This implies that the covering fraction during the dip increased to $>97$\%, consistent with a fully covering absorber. 
In contrast the non-dip spectrum is still absorbed, but with a lower covering fraction of $f=0.83\pm0.02$, consistent with the mean OBS\,1 
parameters. Note that the column density and ionization are consistent with the mean OBS\,1 spectrum, with $N_{\rm H}=6.2\pm0.5\times10^{22}$\,cm$^{-2}$ and $\log\xi=3.0\pm0.1$. 

We also tested whether the increase in absorption could be due to a change in column rather than covering. In this case, the covering fraction (with $f=0.87\pm0.02$) was kept tied, while the soft X-ray column was allowed to vary and the overall continuum flux was allowed to adjust (via a constant multiplicative factor)
to within $\pm20$\% of the mean value. 
In this case, the inferred column increased from $N_{\rm H}=5.0\pm0.3\times10^{22}$\,cm$^{-2}$ to $N_{\rm H}=9.1\pm0.7\times10^{22}$\,cm$^{-2}$ during the dip. However, the fit statistic is considerably worse, by $\Delta\chi^2=+44.4$ (for $\Delta \nu=1$), compared to the variable covering case above and the model is not fully able to reproduce the strong absorption trough above 1\,keV in the dip spectrum. Thus the spectral variability could be due to an increase in the covering fraction of the soft X-ray absorber, reaching a maximum full covering during the flux minimum in OBS\,1, while the covering fraction is at its lowest during the brighter OBS\,2 sequence.

To test the variability in a more model independent way, we examined the {\it NuSTAR} count rate lightcurve across the dip portion of the observations, 
comparing the softer 3--5\,keV {\it NuSTAR} band with the continuum dominated 20--40\,keV hard X-ray band. Any absorption variability should preferentially affect the softer band more than the harder band, while if the dip is due to (colorless) changes in the power-law continuum, then both bands should be variable. Figure~\ref{fig:dip} (right panel) shows the lightcurves in these bands, over the first portion of the {\it NuSTAR} observation. The dip is clearly apparent in the softer 3--5\,keV band, occurring 80\,ks into the {\it NuSTAR} observation, consistent with Figure\,1. However no dip or pronounced variability is seen in the 20--40\,keV band at these times. Thus the dip is primarily due to an increase in obscuration, rather than a simple decrease in the intrinsic power-law continuum, as the latter should also produce a decline in flux in the 20--40\,keV band.

\subsection{The Soft X-ray Spectrum} 

We now investigate the soft X-ray spectrum of PDS\,456 in more detail. We first used the RGS grating spectra obtained from both 
{\it XMM-Newton} observations, over the 6--30\,\AA\ range.  
The signal to noise is more limited in the individual RGS spectra of each sequence and as they are not sensitive to any subtle changes in the absorption parameters, we subsequently combined these into a single time-averaged RGS spectrum from both observations. The resulting spectrum was then binned into wavelength bins of width $\Delta\lambda=0.1$\,\AA, approximately the FWHM resolution of the RGS gratings.
Note that the observed 0.5--2.0\,keV flux obtained from the RGS from these observations is $1.0\times10^{-12}$\,erg\,cm$^{-2}$\,s$^{-1}$; this compares to the flux range of $F_{\rm 0.5-2\,keV}=1.6-4.3\times10^{-12}$\,erg\,cm$^{-2}$\,s$^{-1}$ 
for the previous archival {\it XMM-Newton} RGS observations from 2001--2014 \citep{Reeves16}. 
The 2017 soft X-ray spectrum thus represents a historical low flux, where PDS\,456 is highly absorbed and likely dominated 
by any reprocessed components associated with the wind.

A fit to a simple soft power-law, of $\Gamma=2.3\pm0.1$, absorbed only by the neutral Galactic column, yields an extremely poor fit to the mean spectrum with $\chi_{\nu}^{2}=358/229$, rejected with a null hypothesis probability of $9.6\times10^{-8}$. Similar residuals are seen in the soft X-ray band, as per the EPIC-pn CCD spectrum, 
with a broad absorption trough between 10--12\,\AA\ and an excess of emission between 12--17\,\AA.
Thus we adopt a similar continuum parameterization for the RGS as per the broad-band spectrum above, where the power-law continuum is partially covered by the soft band absorber (and where a fraction $1-f$ remains unattenuated).
A broadened emission component is also included for the excess soft X-ray emission, as per the above broad-band analysis.
The photon index of the intrinsic continuum was fixed at $\Gamma=2.3$, as was determined by the pn and {\it NuSTAR} spectra, 
as otherwise this cannot be determined over the soft X-ray band below 2\,keV, where little of the primary power-law emission 
is seen in this highly absorbed spectrum.

\begin{figure*}
\begin{center}
\rotatebox{-90}{\includegraphics[height=13cm]{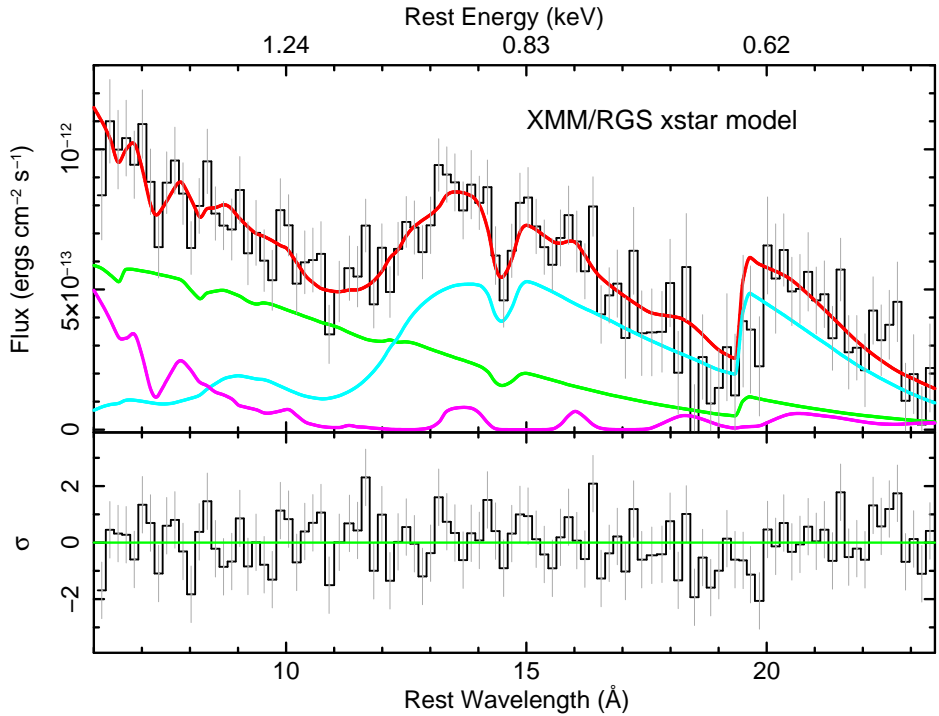}}
\end{center}
\caption{\small The mean {\it XMM-Newton} RGS\,1+2 spectrum of PDS\,456, combined for both OBS\,1 and OBS\,2 to maximize signal to noise. Data points are shown as black histograms (with grey error bars), while the best fit \textsc{xstar} model is overlaid in red (solid line). The model components (solid lines) are as per Figure\,5. 
Several features are apparent in the spectrum. A strong, broad excess of emission is observed between $12-17$\,\AA, which can be modeled by a velocity broadened emission component from a wide angle wind (cyan curve). The broad trough between $9-12$\,\AA\ arises from the low ionization absorption component (zone 2, magneta curve) and is likely produced by a blend of iron L-shell transitions.
A significant absorption line is also apparent at 14.5\,\AA. If this is associated with blue-shifted O\,\textsc{viii} Ly$\alpha$ (at 19.0\,\AA), then its outflow velocity of $-0.25c$ is consistent with the iron K absorber. Note the drop near 20\,\AA\ (23\,\AA\ observed frame), 
arises from the neutral O\,\textsc{i} edge, associated with absorption from our Galaxy. } 
\label{fig:rgs}
\end{figure*}

The mean RGS spectrum, with the best-fit model and components overlaid, is shown in Figure~\ref{fig:rgs}. The soft absorber strongly attenuates the absorbed power-law (Figure~\ref{fig:rgs}, magenta line), which only emerges 
short-wards of 10\,\AA\ and is almost fully opaque at longer wavelengths, while the unattenuated power-law component (green line) is accounted for in the partial covering model. The parameters of the soft X-ray absorber are similar to what was found for the broad-band spectrum; a column of $N_{\rm H}=6.0^{+2.3}_{-2.7}\times10^{22}$\,cm$^{-2}$, an ionization parameter of $\log\xi=3.0\pm0.4$ and a line of sight covering fraction of $f=0.75^{+0.06}_{-0.09}$.

The broadened \textsc{xstar} emission component (cyan line) then accounts for the excess emission between 12--17\,\AA\ and dominates the spectrum at long wavelengths. Note this emitter has a large velocity broadening to account for this strong excess, 
with a Gaussian velocity width of $\sigma_{\rm v}=30000^{+12000}_{-6000}$\,km\,s$^{-1}$. The column density of the emitter 
is assumed to be equal to the soft X-ray absorber, where $N_{\rm H}=6.0\times10^{22}$\,cm$^{-2}$ as above, as this dominates the 
opacity in this band. Thus the implicit assumption in the model is that the absorbing and emitting gas are 
essentially the same, with the latter integrated along all sight lines and the former only seen along the direct line of sight. 
This appears to be supported by the ionization parameter of the emitter, 
which is $\log\xi=2.74\pm0.12$ and is consistent within errors with the ionization of the soft absorber.
Note the broad excess can also be parameterized with a simple Gaussian function, with a width of $\sigma=90^{+40}_{-40}$\,eV, at an energy centroid of $837\pm17$\,eV. The addition of the emission component significantly improved the fit by $\Delta\chi^2=79$, 
compared to an absorbed power-law continuum. The best fit parameters of the RGS model are tabulated in Table\,4.

A narrower absorption feature also appears to be superimposed on the broadened emission, at a rest frame wavelength of 
$\lambda=14.81\pm0.09$\,\AA\ (or $E=837\pm5$\,eV) and an equivalent width of $-9.5\pm4.0$\,eV and its addition further 
improved the fit statistic further by $\Delta\chi^{2}=23$. If this feature is associated with the strong blueshifted O\,\textsc{viii} Ly$\alpha$ doublet at 18.97\,\AA\ (or 653.5\,eV), then this would imply an outflow velocity of $v/c=-0.242\pm0.009$, which would be consistent with the velocity of $-0.25c$ obtained from the Fe K absorber (zone 1a). Indeed this is the strongest soft X-ray absorption line predicted to be present from this zone of gas, as is seen in Figure~\ref{fig:transmission}. An identification with a slower (or stationary) absorption line appears less likely, the nearest absorption line at $\lambda_{\rm rest}=14.82$\,\AA\ occurs from O\,\textsc{viii} $1s\rightarrow5p$, which one would expect to be much weaker compared to the corresponding $1s\rightarrow2p$ line. 
Thus an identification with O\,\textsc{viii} Ly$\alpha$, 
at a self consistent velocity as per the Fe K absorber, seems the most likely interpretation.

To further test this, we fitted for a high ionization \textsc{xstar} component of the absorber, fully covering the RGS spectrum, 
in order to account for the possible O\,\textsc{viii} Ly$\alpha$ absorption. Indeed the parameters are consistent 
with the zone 1a absorber above, with; $N_{\rm H}=2.8^{+2.0}_{-1.5}\times10^{23}$\,cm$^{-2}$, $\log\xi=5.3^{+0.5}_{-0.2}$ and 
$v/c=-0.260\pm0.008$. This suggests that the O\,\textsc{viii} absorption originates as part of the same outflow measured at iron K.
The final fit statistic, with all these components added (two absorbers plus a soft emitter), is then acceptable, with 
$\chi_{\nu}^{2}=209/218$.

\begin{deluxetable}{lc}
\tablecaption{Parameters for the 2017 RGS Spectrum.}
\tablewidth{0pt}
\tablehead{\colhead{Parameter} & \colhead{value}}
\startdata
High ionization absorber, zone 1a:-\\
$N_{\rm H}$$^{a}$ & $2.8^{+2.0}_{-1.5}$ \\
$\log\xi$$^{b}$ & $5.3^{+0.5}_{-0.2}$ \\
$v/c$  & $-0.260\pm0.008$ \\
\hline
Soft X-ray absorber, zone 2:-\\
$N_{\rm H}$$^{a}$ & $0.60^{+0.23}_{-0.27}$ \\
$\log\xi$$^{b}$ & $3.0\pm0.4$ \\
$v/c$  & $-0.23\pm0.02$ \\
$f^c$ & $0.75^{+0.06}_{-0.11}$ \\
\hline
Soft X-ray Emission:-\\
$N_{\rm H}$$^{a}$ & $0.60^f$ \\
$\log\xi$$^{b}$ & $2.74^{+0.12}_{-0.10}$ \\
$\kappa_{\rm xstar}$$^{d}$ & $1.03\pm0.11\times10^{-3}$ \\
$\Omega/4\pi$$^{e}$ & $1.1\pm0.1$ \\
\hline
Power-law Continuum:-\\
$\Gamma$ & $2.3^{f}$ \\
$N_{\rm ABS}$$^{g}$ & $0.55\pm0.07$ \\
$N_{\rm UNABS}$$^{g}$ & $1.6\pm0.5$ \\ 
$F_{\rm 0.5-2.0\,keV}$$^{h}$ & 1.02 \\
\enddata
\tablenotetext{a}{Units of column density $\times10^{23}$\,cm$^{-2}$.}
\tablenotetext{b}{Ionization parameter in units of erg\,cm\,s$^{-1}$.}
\tablenotetext{c}{Covering fraction of absorber.}
\tablenotetext{d}{Measured normalization of \textsc{xstar} emission component, as described in Table\,3.}
\tablenotetext{e}{Solid angle of the emission component, as per Table\,3.}
\tablenotetext{f}{Denotes parameter is fixed.}
\tablenotetext{g}{Relative normalizations of the absorbed and unabsorbed power-law continuum components. Normalizations are in 
flux units of $\times10^{-3}$\,photons\,cm$^{-2}$\,s$^{-1}$\,keV$^{-1}$ at 1\,keV.}
\tablenotetext{h}{Observed 0.5--2\,keV flux, not corrected for absorption, in units of $\times10^{-12}$\,erg\,cm$^{-2}$\,s$^{-1}$.}
\tablenotetext{t}{Denotes parameter is tied between observations.}
\label{tab:rgs}
\end{deluxetable}

\subsection{The Soft X-ray Emission}

The flux of the soft X-ray emission can also be used to infer 
the global covering factor of the outflowing gas; e.g. see \citet{Reeves18b}, for a similar calculation 
in PG\,1211+143. From the photoionization modeling, 
the normalization (or flux), $\kappa$, of an emission component is defined 
by {\sc xstar} in terms of:
\begin{equation}
\kappa = f_{\rm cov}\frac{L_{38}}{D_{\rm kpc}^2}
\end{equation}
where $L_{38}$ is the ionizing luminosity over the 1--1000\,Rydberg band in units of $10^{38}$\,erg\,s$^{-1}$ and 
$D_{\rm kpc}$ is the distance to the quasar in kpc. Here $f_{\rm cov}$ is the covering fraction 
of the gas with respect to the total solid angle, where $f_{\rm cov} = \Omega / 4\pi$. For a 
spherical shell of gas, $f_{\rm cov}=1$, while $L$ is the quasar luminosity that illuminates 
the photoionized shell. Thus by comparing the predicted normalisation ($\kappa$) for 
a fully covering shell of gas illuminated by a luminosity $L$ versus the observed 
normalization ($\kappa_{\rm xstar}$, as tabulated in Tables 3 and 4) determined from the photoionization modeling, 
the global covering fraction of the gas can be estimated. For PDS\,456, the ionizing luminosity is estimated to be
$L=5\times10^{46}$\,erg\,s$^{-1}$ \citep{Nardini15},  
while its distance is $D=726$\,Mpc. Thus for a spherical shell of emitting gas 
the expected {\sc xstar} normalization is $\kappa=9.5\times10^{-4}$. 
As noted above, the global column density of the emitter is assumed to be equal to that of the soft absorber. 

Compared to the observed normalization factors reported in Table\,4 from the RGS analysis, of 
$\kappa=1.03\pm0.11\times10^{-3}$, the global covering fraction of the soft X-ray emitting gas is estimated to be $f=1.1\pm0.1$, 
consistent with the gas covering a substantial fraction of $4\pi$ steradian.
Note that similar estimates for the global covering fraction are also found for the broad band analysis 
from the simultaneous pn and {\it NuSTAR} spectra (see Table\,3). 
This is consistent with the result obtained by \citet{Nardini15} 
for the high ionization wind measured at Fe K, which was found to cover at least 
$2\pi$\,steradian solid angle. Similar results were also obtained by \citet{Reeves16} 
from considering the soft X-ray spectra of the archival {\it XMM-Newton} observations of PDS\,456.

A more intuitive estimate can be made by comparing the luminosity of the soft X-ray emission component versus 
the continuum luminosity which is absorbed by the soft X-ray absorber. If the gas subtends a solid angle close to $4\pi$\,sr and the continuum emission is isotropic, the above ratio will be close to one, as most of the radiation that is absorbed will be subsequently re-emitted. 
Using the above spectral fit for the broad-band {\it NuSTAR} and pn data, then the X-ray luminosity of the emitter is $L_{\rm 0.3-10\,keV}=1.5\times10^{44}$\,erg\,s$^{-1}$, 
while the luminosity absorbed is $L_{\rm 0.3-10\,keV}=4.7\times10^{44}$\,erg\,s$^{-1}$ and thus the above ratio is $\sim 0.3$. Thus approximately 30\% of the absorbed continuum emission is subsequently re-emitted in the soft X-ray band, which is consistent with a relatively high global covering fraction of the gas.
The soft X-ray emission component dominates in these low flux observations due to the intrinsic continuum being highly absorbed, which revealed the reprocessed components originating from the wind. 

As noted above, the velocity broadening of the emission was found to be very high, 
with $\sigma_{\rm v}=30000$\,km\,s$^{-1}$, as can be seen from the broad 
emission present below 1\,keV. Note that by itself, no net velocity shift is required for the emitter. 
However the velocity width is consistent with the observer viewing both the red and blue-shifted 
emission integrated across all lines of sight towards a wide angle expanding wind. Indeed the FWHM velocity of the emitter, 
which would arise across all sight-lines, is similar in magnitude to the net outflow velocity of $0.25c$ that is viewed directly along the line of sight. This may suggest that the soft X-ray emission and absorption arise from the same wide-angle wind.

\section{Multiwavelength Observations of PDS 456}

\subsection{Simultaneous HST Observation}

Simultaneously with the \textit{XMM-Newton/NuSTAR} observations, a UV spectrum of PDS\,456 was also acquired with the Cosmic Origins Spectrograph (COS) onboard \textit{HST}, using the G140L grating. Note this was the first {\it simultaneous} UV vs X-ray spectrum of PDS\,456, 
where the earlier archival observations taken with HST/STIS in 2000 \citep{O'Brien05} and 
with HST/COS in 2014 \citep{Hamann18} were not concurrent with any of the past X-ray observations.
The data were retrieved from the Mikulski Archive for Space Telescopes with no further reprocessing. The \textit{HST} exposure fell astride the beginning of the first \textit{XMM-Newton} observation (OBS\,1), for a total net time of 4643\,s. Following \citet{Hamann18}, we corrected for Galactic reddening by adopting the extinction law from \citet{Cardelli89} with $R_V = 3.1$ and $E(B-V) = 0.45$. The continuum in 2017 was then fitted as a power law over line-free wavelength intervals, returning an index of $\alpha_\lambda = -1.70$, virtually coincident with the value estimated for the 2000 observation ($-1.68$). Indeed, from a simple visual inspection, the 2000, 2014 and 2017 \textit{HST} spectra of PDS\,456 have a remarkably similar continuum shape, once the different flux levels are taken into account. There are some modest variations in the continuum flux, where the earliest 2000 spectrum shows the highest flux recorded so far, which had dropped by about 1/3 by 2014 before slightly recovering to about 4/5 of the original flux in 2017. 

\begin{figure}
\begin{center}
\rotatebox{0}{\includegraphics[width=8.7cm]{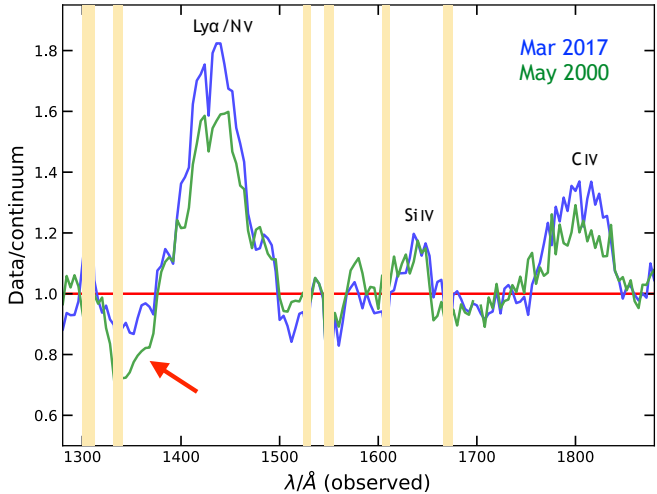}}
\end{center}
\caption{\small Comparison between the 2000 and 2017 UV epochs of PDS\,456 observed with HST. 
Note the narrow yellow bands mark Galactic features. 
In the 2000 HST/STIS spectrum, a broad absorption trough (red arrow) 
is observed at $\sim1350$\AA\ (1140\AA\ rest frame), bluewards of Ly$\alpha$. 
The BAL feature may be identified with Ly$\alpha$ or N\,\textsc{v} shifted by $0.06-0.08c$, or with a C\,\textsc{iv} BAL at $\sim0.3c$ 
\citep{Hamann18}. The BAL feature has substantially 
weakened in March 2017, despite the X-rays being highly absorbed.}
\label{fig:hst}
\end{figure}

The 2017 {\it HST}/COS versus 2000 {\it HST}/STIS spectra are shown in Figure~\ref{fig:hst},
where the de-reddened spectra have been plotted as a ratio against the best fit continuum slope described above. 
The broad emission line profiles from Ly$\alpha$ (also blended with N\,\textsc{v}), C\,\textsc{iv} and Si\,\textsc{iv} are consistent between epochs, 
with the lines having a marginally higher equivalent width in the 2017 epoch, compared to 2000. 
Note that the emission line profiles from the 2000 and 2014 epochs are discussed in detail in \citet{O'Brien05} and \citet{Hamann18}, where in 
particular the C\,\textsc{iv} emission profile is blueshifted by $-5000$\,km\,s$^{-1}$.

In the context of this paper, it is worth checking the spectra for the presence of any BAL-like features, in order to gain further insights on the physical relation between the (soft) X-ray and UV absorbers. The latter has manifested itself in the form of a broad absorption trough at 1346\,\AA~(observed) in the \textit{HST}/STIS spectrum taken in 2000, first revealed by \citet{O'Brien05} and recently identified as a possible C\,\textsc{iv} line outflowing at $\sim$0.3$c$ by \citet{Hamann18}. In agreement with the 2014 \textit{HST}/COS spectrum, there is no evidence for a prominent trough in the 2017 data shortwards of the Ly$\alpha$ emission line, which only appears to be present in the 2000 spectrum, but just a hint of a shallow depression.  Figure~\ref{fig:hst} shows a comparison over the band of interest between the 2000 and 2017 spectra, normalized to the continuum level and visually rebinned for clarity. We included a Gaussian profile to reproduce any possible absorption feature at the position of the putative C\,\textsc{iv} blueshifted line. We obtained a central (observed) wavelength $\lambda_{\rm o} = 1339.1 \pm 1.5$\,\AA~(against $1346.1 \pm 0.6$\,\AA\ for the 2000 feature; see \citet{Hamann18}), optical depth $\tau_{\rm o} = 0.14 \pm 0.01$ (against $0.35 \pm 0.01$), and a Doppler broadening of $b = 6321 \pm 718$ km s$^{-1}$ (against $5135 \pm 203$ km s$^{-1}$). We stress, however that in the 2017 
spectrum, the properties of the very shallow trough heavily depend on the exact level (and slope) of the continuum. Thus we do not consider the detection of this feature as fully reliable compared to the deeper 2000 trough and we take its optical depth as an indicative upper limit.

\subsection{The Optical to X-ray Spectral Energy Distribution}

We constructed the SED of PDS 456 during the 2017 observations, also using the data from 
OBS\,1 during the minimum in X-ray flux. It may be the case that the lower ionizing 
flux during the current observations affects the overall 
properties of the absorber and thus it is instructive to compare the 2017 SED with that obtained during previous observations.  
The 2017 SED was constructed using the same phenomenological form as in \citet{Nardini15} and \citet{Matzeu16}, where a 
double broken power-law form was used for the continuum from the optical to hard X-ray band. Simultaneous photometry from the {\it XMM-Newton} Optical Monitor was used, in the V, B, U, UVW1, UVW2 and UVM2 bands, as well as simultaneous photometry taken with the V and UVW1 filters from the {\it Swift} UVOT, which was also coincident with OBS\,1. The optical/UV data-points were corrected for an reddening as per the HST observations.
Both the soft X-ray and the iron K absorption components are included in the model, as parameterized above in Table~3, as 
well as the Galactic absorption.   
The underlying continuum shape can then be described by three photon indices; $\Gamma=1.1\pm0.1$ in the optical/UV band up to a break energy of 10\,eV, a steep photon index of $\Gamma=3.33\pm0.05$ connecting the UV to the soft X-ray band up to a break energy of 500\,eV and a photon index of $\Gamma=2.34\pm0.06$ from 0.5--40\,keV in the X-rays.

\begin{figure*}
\begin{center}
\rotatebox{-90}{\includegraphics[width=8.7cm]{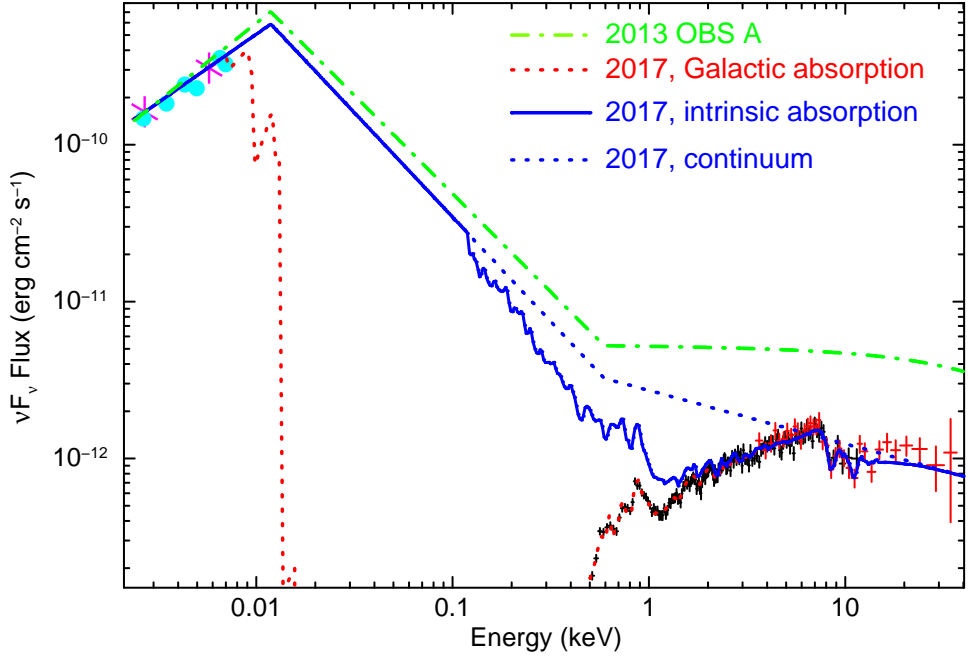}}
\end{center}
\caption{\small The 2017 optical/UV to X-ray SED of PDS\,456. Data points are taken from the OBS\,1 sequence, consisting of 
{\it XMM-Newton} pn (black crosses), {\it NuSTAR} (red crosses), {\it Swift} UVOT (magenta crosses) and {\it XMM-Newton} OM (cyan circles). 
The optical/UV photometric points are corrected for reddening. 
The figure shows various models fitted to the SED, using a double broken power-law continuum. The red dotted line 
shows the SED model, absorbed by the Galactic column as well as by intrinsic absorption from the wind. The blue solid line 
shows the model corrected for Galactic absorption, but still attenuated by the wind, while the dashed blue line shows the 
intrinsic continuum after all absorption is removed. Note the wind mainly attenuates the soft X-ray band. 
In contrast the green dot-dashed line shows the intrinsic continuum from the 
brighter, 2013 OBS\,A {\it XMM-Newton} and {\it NuSTAR} observation, which was largely unabsorbed by the wind; see \citet{Nardini15} for details. The 2017 observation is a factor of $\times4$ fainter in the X-ray band compared to 2013, although there is little difference 
in the optical/UV.}
\label{fig:sed}
\end{figure*}

The 2017 SED is shown in Figure~\ref{fig:sed}, where the emission peaks in the UV band compared to the X-ray. 
Overall the 1--1000\,Rydberg ionizing luminosity of this SED is $L=3.5\times10^{46}$\,erg\,s$^{-1}$, corrected for Galactic absorption and reddening, but not for any intrinsic absorption from the wind. The effect of the X-ray absorption is apparent in the Figure, where the dotted blue line shows the intrinsic continuum level and the solid line the continuum after absorption through the wind.
The primary attenuation occurs in the soft X-ray band, where a maximum in absorption opacity occurs near to 1\,keV. However the bolometric output is dominated by the optical/UV band, which as discussed above, appears relatively unaffected by the presence of the X-ray absorber.

The 2017 SED was also compared with what was computed by \citet{Nardini15}, on the basis of the 2013--2014 {\it XMM-Newton} and {\it NuSTAR} observations. We reconstructed the SED from the first sequence of that campaign (OBS\,A), which was also the brightest and least absorbed of those five earlier sequences; the SED model of the 2013 observation is shown in Figure~\ref{fig:sed} (green dot-dashed line). 
The 1-1000\,Rydberg luminosity is only slightly higher in 2013 OBS\,A, with $L=5.0\times10^{46}$\,erg\,s$^{-1}$. The main difference between 2013 OBS\,A and the 2017 SEDs lies in the X-ray band, where the observed 0.3-30\,keV band luminosity has 
decreased by a factor of $\times4$, from $L_{\rm X}=2.0\times10^{45}$\,erg\,s$^{-1}$ to $L_{\rm X}=5.7\times10^{44}$\,erg\,s$^{-1}$. Thus, while the X-ray luminosity has decreased between these observations, the optical/UV portion remained unchanged.
Indeed, as will be discussed in a subsequent paper on the long term X-ray vs UV variability from the 2017 {\it Swift} campaign, most of the pronounced variability occurs in the X-ray band, with relatively little ($<10$\%) change in the UV flux over a period of 6 months. We note that a similar behavior can occur in other AGN, for instance in the Seyfert 1 Mrk 335, where large amplitude X-ray variability is present, but the UV variability is less pronounced \citep{Gallo18}.

\section{Discussion} \label{sec:discussion}

\subsection{Properties of The Soft X-ray Absorber}

We now discuss the possible location and properties of the soft X-ray wind in PDS\,456, compared to the well studied outflow that 
is present at iron K. 
The previous campaigns on PDS\,456 were able to place constraints on the location of the iron K outflow, based on its variability. 
\citet{Nardini15} estimated a likely radius of $\sim200R_{\rm g}$\footnote{Note that in PDS\,456, with a likely black hole mass of $10^{9}~{\rm M}_{\odot}$ 
(Reeves et al. 2014), $1R_{\rm g}=GM/c^2 = 1.5\times10^{14}$\,cm} 
for the $0.25c$ Fe K absorber. This was based on the typical response time, of one week,  
of the P Cygni like profile at Fe K to changes in the hard X-ray continuum, as measured during the series of five simultaneous {\it XMM-Newton} and {\it NuSTAR} observations in 2013--2014. Rapid variability of the iron K absorber was also observed during the 2013 {\it Suzaku} observations of 
PDS\,456, which covered a baseline of $\sim1$\,Ms \citep{Gofford14,Matzeu16}. 
PDS\,456 had a low X-ray flux during those observations and the iron K absorber increased by a factor of $\times 10$ in 
equivalent width during the {\it Suzaku} campaign. Based on this variability, \citet{Matzeu16} estimated a distance of $R\sim200R_{\rm g}=3\times10^{16}$\,cm, assuming that the variability was due to an increase
in the column density of the outflow.  A slightly larger radial distance estimate was obtained by \citet{Gofford14}, of 
about $R\sim1000R_{\rm g}\sim10^{17}$\,cm, from the recombination timescale, on the assumption that the ionization of the iron K absorber responded to the decrease in the X-ray continuum flux during those observations. 

An interesting question is what is the location of the lower ionization soft X-ray absorber, relative to the iron K absorber? Adopting the above distance of $R=3\times10^{16}$\,cm for the iron K wind, then for an ionization of $\log \xi=5$ (Table~\ref{tab:xstar}, zone 1a) and an ionizing luminosity of $3\times10^{46}$\,erg\,s$^{-1}$ (Section 4.2), 
its density is $n_{\rm e}=L/\xi R^{2} \sim 3\times10^{8}$\,cm$^{-3}$. 
In contrast, the soft X-ray absorber is $\times100$ 
less ionized, with $\log\xi=3$ and so this would require its density to be $\times100$ higher to compensate, i.e. 
$n_{\rm e} \sim 3\times10^{10}$\,cm$^{-3}$. Given the soft X-ray column density of $N_{\rm H}=6\times10^{22}$\,cm$^{-2}$, then 
the absorber sizescale is $\Delta R \sim N_{\rm H} / n_{\rm e} = 2\times10^{12}$\,cm.
This seems less likely, as it implies that the soft X-ray absorber exists in the form of many dense but 
extremely small clumps, of $<<1R_{\rm g}$ in size, so as to not become over ionized close to the X-ray source.

Instead the soft X-ray absorber could be located further out than the high ionization zones of the wind, where a 
factor of $\times10$ increase in distance would translate to a decrease in ionization of $\times100$ for a given density.
We can place an estimate on the radial location of the soft X-ray absorber in PDS\,456 from its variability between 
OBS\,1 and OBS\,2.
In particular, during the dip in OBS\,1, the soft X-ray absorber appeared to reach a maximum covering of 100\% of the X-ray source, 
which then decreased to about 70\% during OBS\,2 as the AGN brightened. Thus we likely witnessed an occultation event, whereby the 
absorbing clump or cloud fully covered the X-ray source during the dip and then gradually moved out of the line of sight as the X-ray observations progressed.

One possible assumption is that the absorber size scale ($\Delta R$) is approximately equal to the X-ray source size ($D$).
If the absorbing cloud were much smaller, then it would not be able to cover such a high fraction (up to 100\%) of the X-ray source, while if it were substantially larger, the variations in covering fraction would likely not be seen.
The X-ray source or coronal size can be estimated from the light crossing time, i.e. $D=c\Delta t$, where $\Delta t$ corresponds to the doubling time of the 
X-ray flares. Figures\,1 and 2 shows various rapid X-ray flares which double in flux on timescales of 10s of kiloseconds,
for instance at 280\,ks into the {\it NuSTAR} observation or after 3\,days into the {\it Swift} monitoring. Indeed, such rapid X-ray flaring has been noted previously in PDS\,456, from {\it RXTE}, {\it Beppo-SAX} and 
{\it Suzaku} observations \citep{Reeves00, Reeves02, Matzeu16} and implies that the X-ray source is a few gravitational radii 
in extent.

If the obscuring cloud has a sizescale $\Delta R \approx D = c\Delta t$, with column density $N_{\rm H} = n_{\rm H} \Delta R$ (where the electron density $n_{\rm e} \sim n_{\rm H}$) and ionization parameter $\xi = L_{\rm ion}/n_{\rm e} R^{2}$, then its distance, $R$, is:-
\begin{equation}
R \approx \left( \frac{L_{\rm ion} c \Delta t}{N_{\rm H} \xi} \right)^{1/2} 
\end{equation}
\noindent From the X-ray spectral analysis, the column density of the soft X-ray absorber is $N_{\rm H}=6\times10^{22}$\,cm$^{-2}$, while its 
ionization is $\log\xi=3$ (see Table~\ref{tab:xstar}). 
The ionizing luminosity, as defined by \textsc{xstar} over the 1--1000\,Rydberg range, 
is $L=3\times10^{46}$\,erg\,s$^{-1}$, measured from the SED analysis in Section\,4.2.

The doubling time is estimated to be $\Delta t=20$\,ks from the X-ray flares and thus $D=6\times10^{14}$\,cm (or $\sim 10^{15}$\,cm).
The above values yield $R=5\times10^{17}$\,cm or $\sim0.1$\,pc. This is consistent with a similar estimate in \citet{Reeves16}, 
based on an analysis of the soft X-ray absorber variations over the previous {\it XMM-Newton} observations of PDS\,456 from 
2001--2014. 
The larger radial distance to the soft X-ray gas, compared to the high ionization absorber, likely prevents it from becoming 
over ionized.
At this distance the Keplerian velocity is $\sim5000$\,km\,s$^{-1}$, while the gas density is $n_{\rm e}\sim 10^{8}$\,cm$^{-3}$, consistent 
with a BLR scale origin.

The transverse velocity of the absorbing clouds can also be estimated, using the time taken for the X-ray source to uncover. 
In Section\,3.3, we determined that the absorber covering fraction varied from $f>0.97$ during the dip in OBS\,1, to $f=0.83\pm0.02$ 
in the remainder of OBS\,1 and then declining further to $f=0.73\pm0.03$ in OBS\,2. Thus the change in covering fraction is $\Delta f\sim0.25$, 
over a baseline of $\Delta t_{\rm uncov} = 150$\,ks (i.e. one {\it XMM-Newton} orbit). The cloud transverse velocity is then:-

\begin{equation}
v_{\rm t} \approx \frac{\Delta f D}{\Delta t _{\rm uncov}} 
\end{equation}

\noindent This gives a transverse velocity of $v_{\rm t}\sim10^{4}$\,km\,s$^{-1}$. This is within a factor of two of the Keplerian velocity, while the wind may also have a transverse component across the line of sight.
These transiting wind clumps may provide an explanation for the variable partial covering previously seen in the 2013 {\it Suzaku} observations of PDS\,456, \citep{Matzeu16}, where the QSO was in a similar low flux state and substantial spectral variability was also observed.

A similar scenario was also recently suggested for the fast outflow in the 
quasar PG\,1211+143 \citep{Reeves18b}. Here, the fast wind consists of a two phase (low and high ionization) clumpy medium, where the smaller, denser low ionization clumps may account for the rapid variations in the soft X-ray absorbing gas on timescales of days.
Such a scenario may also be required in BAL quasars, in order to explain the co-existence of both the UV and X-ray absorbing gas \citep{Hamann13}.
Recently, a rapid ($\Delta t=100$\,ks) and large ($\Delta N_{\rm H}=10^{24}$\,cm$^{-2}$) variation 
in the absorbing column was measured in the fast ($\sim0.1c$) wind in the Seyfert\,2, 
MCG\,--3-58-007, during an X-ray occultation event \citep{Braito18}. 
This was likely due to a denser wind streamline moving across the line of sight, 
which occurred during the course of a {\it NuSTAR} observation. Indeed, similar to the case of PDS\,456, the decline in flux was measured only in the softer 
X-ray bands below 10\,keV, whereas the hard X-ray continuum did not vary during the absorption event.

Overall, it appears likely that the accretion disk winds are not homogeneous 
in nature and the passage of dense, high column density clumps could explain their short timescale variability, as is predicted in time-dependent wind simulations \citep{Proga00, ProgaKallman04}. Furthermore the X-ray eclipse in PDS\,456 may be related to X-ray obscuration events that occur in other AGN \citep{Markowitz14}, but at lower outflow velocities. For example, the Seyfert 1 NGC 3783 has recently dropped into a low X-ray flux state, due to a pronounced increase in X-ray absorption \citep{Mehdipour17} and which may be caused by frequent obscuration events \citep{Kaastra18}. Notably, a long-term (and on-going) increase in obscuration has been seen towards NGC\,5548 \citep{Kaastra14}. A rapid X-ray absorption event has also been recently detected in {\it XMM-Newton} observations of the Seyfert 1, NGC\,3227 \citep{Turner18}, as was evident through a similar pronounced dip in the soft X-ray lightcurve and which was accompanied by an increase in the soft X-ray column or covering fraction. The opposite can also occur, in the form of uncovering events, as was found for example in the classic changing look AGN, NGC\,1365 \citep{Braito14}. In all of these cases, the rapidly variable X-ray obscuration is likely linked with inhomogeneous matter, possibly originating from an accretion disk wind. In PDS\,456, the event is only more extreme due to the large velocity of the outflowing gas.

\subsection{The Ultra Fast Zone}

In contrast to the soft X-ray absorber, the fastest ($\sim0.4c$), highest ionization ($\log\xi=6$) phase of the wind likely originates from closest to the black hole, while the column of this component may be as high as $N_{\rm H}=7\times10^{23}$\,cm$^{-2}$ (OBS\,1, Table\,3, zone 1b). 
An upper-limit to its radial distance arises from the geometric condition that $\Delta R / R<1$, 
which when combined with the electron density of $n_{\rm e}= N_{\rm H}/\Delta R$ and the definition for the ionization parameter, 
yields $R<L/N_{\rm H} \xi$; see also \citet{Tombesi13} or \citet{Gofford15} for similar radial estimates. 
For the above parameters, $R<4\times10^{16}$\,cm, which is consistent with the estimate of $\sim200R_{\rm g}$ in 
\cite{Nardini15} for the $0.25c$ iron K absorber. A minimum distance derives from the escape radius, where $R_{\rm esc}=2(c^2/v^2) R_{\rm g} = 
10R_{\rm g} \sim10^{15}$\,cm. 

While the exact distance of the fastest zone is not well determined, the above estimates are consistent with it being 
an innermost streamline of the wind, which given its velocity, may be launched from close to the black hole.
As was discussed in paper I, this fast phase might only be detected when the incident X-ray flux is 
relatively low, as per these 2017 observations, so that the gas is not fully ionized. The only previous observations of PDS\,456 where the X-ray flux was this low occurred during the long 2013 {\it Suzaku} campaign. However, those observations lacked the sensitivity of 
{\it NuSTAR} above 10\,keV for detecting this putative high velocity wind component. 
Further {\it NuSTAR} observations will test whether the ultra fast zone is only apparent during observations at low X-ray flux.

\subsection{UV versus X-ray Absorption}

In the original 2000 HST/STIS spectrum of PDS\,456 \citep{O'Brien05}, a broad absorption trough was observed at 1346\,\AA\ (1140\,\AA\ rest frame), blue-shifted by $-0.06c$ with respect to the strong broad Ly$\alpha$ emission line (see Figure~\ref{fig:hst}). \citet{Hamann18} recently reanalyzed those data and found that this trough is unlikely to be associated with Ly$\alpha$ due to the very low ionization that is required, without other lines from metal ions being present. 
They suggested that the trough is most likely either associated with N\,\textsc{v} $\lambda1240$\AA, blue-shifted by $-0.08c$, or from 
highly blueshifted C\,\textsc{iv} $\lambda1549$\AA\ at a velocity of $-0.3c$. Note the latter velocity is also commensurate with the 
outflow velocity typically measured in the X-rays from the Fe K profile \citep{Reeves09, Matzeu17}.
Despite the substantial X-ray absorption seen towards PDS\,456 in the simultaneous 2017 observations, the 1346\,\AA\ feature appears to have decreased in depth between 2000 and 2017, where the optical depth of the trough declined from $\tau=0.35$ to $\tau=0.14$ (see Section 4.1). Unlike the X-ray observations, the simultaneous 2017 HST/COS observation of PDS\,456 shows little intrinsic absorption in the UV, while most of the absorption towards PDS\,456 occurs in the soft X-ray band.

The highly ionized component of the wind, with $\log\xi=5$, is far too highly ionized to produce detectable absorption feature in the UV, as most of the absorption is associated with He and H-like iron and only trace amounts of soft X-ray absorption is present, such as from O\,\textsc{viii} Ly$\alpha$. 
However the lower ionization soft X-ray absorber, with $\log\xi=3$, may produce detectable UV absorption from the wind. 
To test this, we ran \textsc{cloudy} \citep{Ferland98} simulations for a single ionized cloud, using the same methods described in \citet{Hamann18}, over a range of possible ionizations and for soft X-ray column densities of $\log(N_{\rm H}/{\rm cm}^{-2})=22.7-23.0$, consistent with what is observed in the 2017 X-ray observations.
The simulations assume a turbulence velocity of $b=5135$\,km\,s$^{-1}$, which was inferred by \citet{Hamann18} from fitting the UV absorption trough in 2000, while Solar metallicity is assumed.
Note that, for the SED of PDS\,456, the predicted conversion between $\log U$ and $\log \xi$ is approximately $\log\xi = \log U + 1.2$. Thus an ionization within \textsc{cloudy} of $\log U=1.8$ is equivalent to the ionization of $\log\xi=3$ inferred by \textsc{xstar} for the soft X-ray absorber.

\begin{figure}
\begin{center}
\rotatebox{0}{\includegraphics[width=8.7cm]{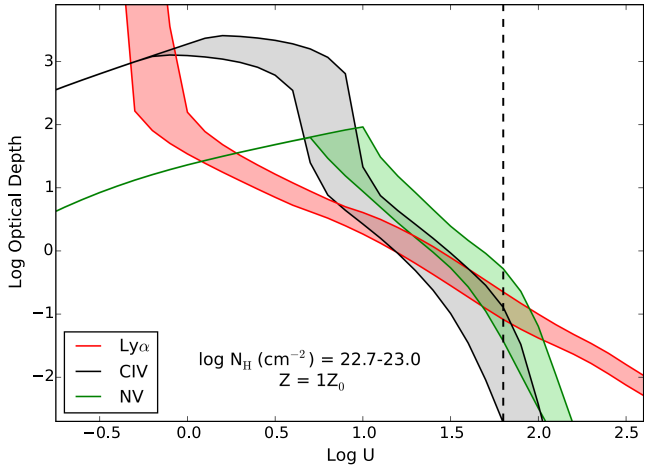}}
\end{center}
\caption{\small \textsc{cloudy} simulations of the UV absorption trough seen in PDS\,456. The simulations predict the optical depth of the 1346\,\AA\ trough, assuming a BAL origin associated with either Ly$\alpha$ (red), C\,\textsc{iv} (black) or N\,\textsc{v} (green), as described in the text. The depth is computed for each ion 
as a function of the ionization ($\log U$) and for Solar metallicity, where the shaded regions correspond to 
column densities in the range between $\log(N_{\rm H}/{\rm cm}^{-2})=22.7-23.0$, as per the observed soft X-ray column. 
The vertical dotted line shows an ionization of $\log U=1.8$, corresponding to the soft X-ray absorber and where the UV absorption trough is predicted to be shallow, where typically $\tau\sim0.1$. Note the optical depths are a very sensitive function of the ionization, thus at much lower ionizations than observed here considerably deeper BAL troughs are predicted.}
\label{fig:cloudy}
\end{figure} 

Figure~\ref{fig:cloudy} shows the predicted optical depth for various lines, such as C\,\textsc{iv}, N\,\textsc{v} or Ly$\alpha$, 
versus ionizaion parameter, $\log U$. Note the shaded area of each curve for a particular ion is bounded by the lower ($\log(N_{\rm H}/{\rm cm}^{-2})=22.7$) and upper ($\log(N_{\rm H}/{\rm cm}^{-2})=23.0$) values of the column density, to highlight the range of uncertainty. 
Thus for the ionization ($\log U=1.8$)  and column expected from the soft X-ray absorber, only small optical depths are predicted for the BAL absorption troughs. 
For C\,\textsc{iv}, the predicted optical depth is $\tau<0.1$ (and similar for N\,\textsc{v}), which is consistent with the 2017 {\it HST} observation where even if any BAL troughs are present, they are very weak and shallow. Thus the soft X-ray absorber observed in 2017 is likely to be too ionized to produce substantial UV absorption and only trace amounts of these ions are present at these ionizations.

However, as can be seen in Figure~\ref{fig:cloudy}, the optical depths inferred from the \textsc{cloudy} model are very sensitive to the ionization.
Indeed the predicted optical depths can increase by more than an order of magnitude, to $\tau>1$, for a decrease in ionization of $\Delta (\log U)=-0.5$. 
Furthermore, the optical depth is also sensitive to the total hydrogen column density. 
If the column density exceeds $N_{\rm H}>10^{23}$\,cm$^{-2}$, then the ionic fractions of both C\,\textsc{iv} and N\,\textsc{v} can increase rapidly, 
due to the effects of shielding within an individual cloud as it becomes optically thick; e.g. see Figure\,8, \citet{Hamann18}. 
Thus only a modest change in column or ionization could explain the appearance (or disappearance) of the UV trough. 
Indeed the soft X-ray column towards PDS\,456 in the highly obscured 2013 {\it Suzaku} 
observations \citep{Matzeu16} were even higher than observed here, exceeding $N_{\rm H}>10^{23}$\,cm$^{-2}$. Unfortunately, no simultaneous UV observations were obtained during the 2013 {\it Suzaku} epoch, while no simultaneous X-ray observations were available from the 2000 
{\it HST} epoch when the deep UV trough was apparent.

It is also possible that the UV observations just missed the X-ray obscuration event in the 2017 observations. 
The {\it HST} observation coincided with the start of the first {\it XMM-Newton} observation, just prior to the onset of the dip in lightcurve in Figure\,1 and thus the subsequent increase in soft X-ray obscuration.
The line of sight towards the UV and X-ray emission regions may also be different. The X-ray emission region is likely located closer to the black hole and thus an outflowing absorbing cloud, which initially passes in front of the central X-ray source, may be delayed in passing across the UV emitter. The UV emitter may also be more extended than the compact X-ray corona and therefore the covering fraction of the absorbing clouds could be much smaller against the UV continuum, which as a result would imprint only very shallow UV absorption troughs.
Indeed, as was inferred earlier in Section 5.1 from the short timescale of the X-ray eclipse event, the absorbing cloud responsible for the increase in X-ray column towards PDS\,456 is likely to be compact, around 10 gravitational radii in extent. Such a small absorbing cloud would have a lower covering fraction compared to the extended optical/UV emission region, where the latter may originate on scales of tens to hundreds of gravitational radii on the accretion disk.
Nonetheless, two further epochs of simultaneous UV and X-ray observations of PDS\,456 have been scheduled with {\it HST}, {\it XMM-Newton} and {\it NuSTAR} over the next year, which will test if there is any link between the UV and soft X-ray absorbers.

\subsection{The Link between UV BALs and X-ray Ultra Fast Outflows}

In general, there appear to be very few examples of ultra fast outflows, discovered in the X-ray band, where there is a corresponding UV counterpart. The nearby ($z=0.0809$) non-BAL QSO PG\,1211+143, which has a well established X-ray ultra fast outflow \citep{Pounds03, PoundsReeves09, Reeves18b}, has recently been found to have a broad UV absorption trough, blueshifted with respect to Ly$\alpha$ by $-17,000$\,km\,s$^{-1}$ \citep{Kriss18a}. This is consistent with the blue-shift (of $-0.06c$) of the soft X-ray absorber, as measured in contemporary X-ray grating observations \citep{Reeves18b, Danehkar18} with {\it XMM-Newton} RGS and {\it Chandra} HETG, as well as with the velocity at iron K \citep{Pounds16}. In PG\,1211+143, the UV mini-BAL trough also appears to be variable and is not detected in all of the epochs analyzed by \citet{Kriss18a}. Thus, aside from PDS\,456 and PG\,1211+143, as of yet there are no other known examples of UV systems which may be associated with X-ray selected ultra fast outflows. Indeed \citet{Kriss18b} perform a follow-on study of 16 archival UV {\it HST} and {\it FUSE} observations of ultra fast outflows selected in the X-ray band and do not confirm any detections of blue-shifted UV absorption troughs.
They suggest that most of the X-ray selected ultra fast outflows are likely to be too highly ionized to have detectable UV absorption features.

However, the reverse is not the case and there are several examples of known UV BAL or mini-BAL systems where ultra fast outflows have subsequently been detected in the X-ray band. The most studied example is the high redshift ($z=3.9$) BAL QSO, APM\,08279+5255, where several observations have clearly established the presence of an ultra fast outflow at iron K with a typical velocity of $0.2-0.4c$ \citep{Chartas02,Chartas09,Saez09,SaezChartas11}, while, similar to PDS\,456, pronounced outflow variability has also been measured. 
Other recent examples include; the BAL QSOs, 
HS~08104+2554 \citep{Chartas14} and H~1413+117 \citep{Chartas07}, as well as the mini-BAL PG\,1115+080 \citep{Chartas03}. These QSO's are all high redshift ($z=1-3$), are gravitationally lensed (boosting their X-ray flux) and have outflow velocities measured at Fe K of between $0.1-0.4c$. 

While the majority of UV BAL systems are detected at high redshift, as their rest-frame UV spectra then fall into the optical band, there are also notable examples of UV BALs/mini-BAL systems at low redshift which also have corresponding X-ray ultra fast outflows. Mrk\,231 is a well known nearby ($z=0.042$) low ionization BAL (LoBAL) QSO, which is also classified as a Ultra Luminous Infra-Red Galaxy \citep{Sanders03}. \citet{Feruglio15} measured both the presence of a fast X-ray wind, with an outflow velocity of $-20,000$\,km\,s$^{-1}$ and a large scale molecular outflow in CO, the latter with a mass outflow rate of up to $\sim1000$\,M$_{\odot}$\,yr$^{-1}$. They also inferred that the large scale outflow was energy conserving, similar to the recent example of the nearby QSO IRAS\,F11119+3257 \citep{Tombesi15}, with the outflow receiving a momentum boost by a factor of several tens; although this boost may in fact be lower \citep{Nardini18}. Another intriguing example is the nearby QSO PG\,1126-041 ($z=0.06$), where a X-ray outflow was detected with a velocity of $-16,500$\,km\,s$^{-1}$, higher than the corresponding UV mini-BAL system \citep{Giustini11}. Similar to PDS\,456, the ionized X-ray absorber was found to be highly variable. 

PDS\,456 appears to show many of the characteristics of the BAL QSOs. The outflow velocity of PDS\,456 is very similar to the high redshift 
BAL QSOs noted above. PDS\,456 is also relatively X-ray quiet and its UV to X-ray spectral index is steep. From the 2017 SED reported above, we find $\alpha_{\rm ox}=2.0\pm0.1$, as measured between the rest frame 2500\,\AA\ and 2\,keV bands \citep{WilkesElvis87}. This is much steeper than the mean $\alpha_{\rm ox}$ measured in radio-quiet QSO's, where for instance \citep{LussoRisaliti16} find a mean value of $<\alpha_{\rm ox}>=1.63\pm0.01$ from their systematic QSO study. 

The steep $\alpha_{\rm ox}$ in PDS\,456 is in line with what is often observed in BAL QSO's. These are known to be X-ray quiet and can often have 
$\alpha_{\rm ox}$ values around 2 or even steeper \citep{Gallagher06}. This may be as a result of their high level of soft X-ray obscuration or from the QSOs being intrinsically X-ray weak. The soft X-ray spectrum observed here is obscured by a column of at least $N_{\rm H}=6\times10^{22}$\,cm$^{-2}$ and thus substantial lower ionization gas is present in the wind in addition to the highly ionized iron K absorption. The steep SED of PDS\,456, as parameterized through its large $\alpha_{\rm ox}$ and from its steep X-ray photon index of $\Gamma=2.3$, may aide in preventing the absorbing gas from becoming overly ionized by a strong hard X-ray continuum. Such a scenario was also recently discussed by 
\citet{Giustini18}, who argue that AGN with high black hole masses accreting at a high fraction of the Eddington limit are most likely to sustain a persistent disk wind. Here, the relative weakness of the X-ray to UV continuum is favorable for supporting a line driven wind and suppressing the ionization of the gas. 

Fast X-ray winds may also be present in other high accretion rate AGN, but with smaller black hole masses, such as in the Narrow Line Seyfert 1s (NLS1s).  
These AGN also have steep X-ray spectra, with weak hard X-ray emission.
Recently, a fast $0.25c$ outflow was detected in the nearby NLS1, IRAS\,13224-3809, from the detections of a blueshifted iron K absorption line \citep{Parker17} and weaker blueshifted lines in the soft X-ray band 
\citep{Pinto18}. Similarly, \citet{Hagino16} were able to model the large spectral drop above 7\,keV in the extreme NLS1 1H\,0707-495 with an accretion disk wind profile, as was originally suggested by \citet{Done07} for this AGN. 
Both of these AGN also show broad blue-shifted high excitation lines in their UV spectra, most notably from 
C\,\textsc{iv} \citep{LeighlyMoore04}, which likely has a wind origin. The C\,\textsc{iv} emission profile of 
PDS\,456 is similar or somewhat larger, where the line centroid is shifted by $-5000$\,km\,s$^{-1}$ when compared to the expected rest wavelength \citep{O'Brien05}. For all of these AGN, a high accretion rate with respect to Eddington is most likely the fundamental parameter for driving 
and sustaining a fast accretion disk wind.

\section{Conclusions}

We presented simultaneous {\it XMM-Newton}, {\it NuSTAR} and {\it HST} observations of PDS\,456, observed in March 2017. The quasar was observed with a low X-ray flux, only matched by an earlier 2013 {\it Suzaku} observation \citep{Matzeu16}, with a high level of X-ray obscuration.
In addition to the fast ($0.25c$, $0.4c$) zones present at iron K, in the {\it XMM-Newton} observations the soft X-ray spectrum was obscured 
by a lower ionization outflowing zone of gas, with a column density of $N_{\rm H}=6\times10^{22}$\,cm$^{-2}$. We found that the covering fraction was variable, reaching a maximum value consistent with 100\% covering during the pronounced dip in the X-ray lightcurve seen in OBS\,1, while the covering then declined as the source brightened thereafter.

Overall, the wind in PDS\,456 is likely to be clumpy or stratified, with the wind ionization ranging by 3 orders of magnitude from the soft X-ray ($\log \xi=3$) up to the fastest iron K zone ($\log\xi=6$).
The soft X-ray gas may be be located further out, at typical distance scales of $\sim0.1$\,pc, compared to the highly ionized iron K absorber, where the latter more likely originates from closer to the black hole nearer to the wind launching point. 
This gas is likely to be clumpy and variations in the absorber covering fraction may explain the spectral variability, both within these observations and with earlier observations of PDS\,456, which were brighter and less absorbed \citep{Reeves09, Nardini15}. 

Potentially the soft X-ray absorber could impart measurable features onto the UV spectrum of PDS\,456, through high excitation lines such as C\,\textsc{iv} or N\,\textsc{v}. However, despite the strong X-ray obscuration, no significant UV absorption features were present in the 2017 {\it HST} spectrum, compared to the strong blueshifted absorption trough present in the original 2000 {\it HST} observation. 
As was discussed in Section\,5.3, this may be due to the relatively high ionization of the soft X-ray absorber in PDS\,456.
Further simultaneous {\it XMM-Newton} and {\it HST} observations will be able to test whether or not the X-ray and UV absorption is concurrent. Nonetheless, PDS\,456 bares many of the characteristics of the classical BAL quasars and its steep SED as well as its high accretion rate are likely to be conducive for supporting a fast accretion disk wind.

\section{Acknowledgements}

JR acknowledges financial support through grants 
NNX17AC38G, NNX17AD56G and HST-GO-14477.001-A. 
VB acknowledges support from the Italian Space Agency
(contracts ASI-INAF  I/037/12/0 and ASI-INAF n.2017-14-H.0)
AL acknowledges support via 
the STFC consolidated grant ST/K001000/1. EN is funded by the EU Horizon 2020 Marie Sklodowska-Curie 
grant no. 664931. 
Based on observations obtained with XMM-Newton, an ESA science mission with instruments and contributions directly funded by ESA Member States and NASA.

\end{document}